\documentclass[]{aa}

    \usepackage{lineno}

    \usepackage{graphicx}
    \usepackage{txfonts}
    \usepackage{booktabs}
    \usepackage[normalem]{ulem}
    \usepackage{multirow}

    \usepackage[colorlinks = true, allcolors=blue]{hyperref}

    \usepackage{stfloats}
    \usepackage{booktabs}
    \usepackage{orcidlink}
    \usepackage{sidecap}
    \usepackage{pdflscape}

    \usepackage[normalem]{ulem}

    \usepackage{xcolor}

\begin{document}

   \title{The origin of UV emission in extremely low-mass satellite  galaxies: Insights from Leo T}

\author{S. Amrutha\orcidlink{0009-0005-6072-9252}\inst{1,2}\fnmsep\thanks{Corresponding author: \email{amrutharao99@gmail.com}}
        \and
        Mousumi Das\orcidlink{0000-0001-8996-6474}\inst{1}\fnmsep\thanks{Corresponding author: \email{mousumi@iiap.res.in}}
        \and
        Daniel Vaz\orcidlink{0000-0002-8119-6812}\inst{3,4}
        \and
        Jarle Brinchmann\orcidlink{0000-0003-4359-8797}\inst{3,5,6,7}
        }

   \institute{Indian Institute of Astrophysics, Koramangala II Block, Bangalore 560034, India\\
         \and
             Pondicherry University, R.V. Nagar, Kalapet, 605014, Puducherry, India\\
              \and Instituto de Astrofísica e Ciências do Espaço, Universidade do Porto, CAUP, Rua das Estrelas, 4150-762 Porto, Portugal \\
         \and
         Centre for Astrophysics and Supercomputing, Swinburne University of Technology, Hawthorn, VIC 3122, Australia\\
         \and European Southern Observatory, Karl-Schwarzschild-Stra.e 2, 85748 Garching, Germany\\
         \and Departamento de Física e Astronomia, Faculdade de Ciências, Universidade do Porto, Rua do Campo Alegre 687, PT4169-007, Porto, Portugal \\
         \and Leiden Observatory, Leiden University, PO Box 9513, 2300 RA Leiden, The Netherlands\\
         }

   \date{Received XXX; accepted XXX}

% \abstract{}{}{}{}{} 
% 5 {} token are mandatory
 
  \abstract
   {The Leo~T Milky Way (MW) satellite is the lowest-mass gas-rich galaxy known to exhibit recent star formation. This highly dark matter-dominated system provides an ideal laboratory for studying star formation in the ultraviolet~(UV) regime within an extreme, low-density, and metal-poor environment.}
  % aims heading (mandatory)
   {We present a far-ultraviolet~(FUV) study of Leo~T using observations from the Ultraviolet Imaging Telescope~(UVIT). Our goals are to characterize the properties of FUV star-forming clusters (SFCs), identify FUV-emitting stars, constrain their physical properties through comparison with existing catalogs, and investigate how star formation proceeds in an extremely low-mass, metal-poor galaxy.}
  % methods heading (mandatory)
   {We examined the FUV extent of the galaxy and compared it with the H\,\textsc{i} distribution. We extracted thirty SFCs and derived their properties, including area, luminosity, and star formation rate~(SFR), and compared them with previous studies. Far-ultraviolet-emitting stars and their stellar parameters were identified through cross-matching with catalogs from the Hubble Space Telescope and the Multi Unit Spectroscopic Explorer (MUSE). We also estimated the global SFR and star formation efficiencies associated with previously identified CO clouds.}
  % results heading (mandatory)
   {The FUV luminosities of the SFCs confirm the absence of massive O- and B-type stars in Leo~T. The FUV-emitting stellar population consists of main-sequence~(MS) stars, blue loop~(BL) stars, horizontal~branch stars, and emission-line~(EM) stars. The upper MS and EM stars are concentrated near the optical center, whereas the lower MS and BL stars are more spatially extended. We find that five out of the six FUV-emitting stars with MUSE spectroscopy are extremely metal-poor ($\mathrm{[Fe/H]}~\leq -3$). The ($\mathrm{NUV - r}$) color profile indicates that the UV emission is dominated by stellar populations younger than 1 Gyr out to a radius $<$~185~pc. Despite the recent detection of CO(1-0) emission, no SFCs are associated with the molecular clouds, which exhibit extremely low star formation efficiencies. The spatial distribution of the SFCs is suggestive of a directional interaction with the MW.}
{}

   \maketitle
   \titlerunning
   \nolinenumbers
%
%
%------------------------------------------------------------------
\section{Introduction}

Milky Way (MW) dwarf satellite galaxies are broadly classified as either gas-rich or gas-poor systems \citep{Jones_2024}. Gas-rich, low-mass satellites typically exhibit ongoing or recent star formation, whereas quenched galaxies are largely devoid of gas. Among these, the faintest satellites are highly compact, making them some of the smallest known galaxies in the Universe \citep{Belokurov_2013}. They are dynamically overwhelmingly dark matter-dominated \citep[e.g.,][]{10.1111/j.1365-2966.2010.16753.x, Walker_2011, 2012AJ....144....4M, 2014MNRAS.437..415D, Brook_2015, Bullock_2017, Read_2019, Orkney_2021}. Chemically, they are extremely metal-poor and are considered to be some of the least evolved stellar systems in the local Universe \citep{2013ApJ...767..134V, 2019ARA&A..57..375S}. These extreme properties make them unique cosmological laboratories for investigating the formation of the first galaxies and for placing stringent constraints on the nature of dark matter. Over the past two decades, the advent of large-scale digital sky surveys has driven a surge in the discovery of these faint systems \citep[see review by][and references therein]{2019ARA&A..57..375S}.

Recent studies of the star formation histories (SFHs) of faint dwarf galaxies, based on deep and accurate Hubble Space Telescope (HST) photometry, have shown that their stellar populations are predominantly old, with a substantial fraction of their stars having formed in the early Universe \citep{2012ApJ...753L..21B, Brown_2014, 2014ApJ...789..147W}. The blue stars observed in these systems have often been attributed to blue stragglers rather than recent star formation \citep{Santana_2013}. These galaxies may therefore be regarded as fossil remnants that preserve information about the early stages of galaxy evolution \citep{2019ARA&A..57..375S}. In addition, several studies have suggested that some faint dwarf galaxies may have shallower initial mass function (IMF) slopes than the canonical Salpeter IMF, corresponding to a bottom-light IMF \citep{2013ApJ...771...29G, Gennaro_2018}. If such an IMF is extrapolated to higher stellar masses, it would imply a relatively larger population of massive stars and hence a higher supernova (SN) rate for a given total stellar mass. The resulting enhanced stellar feedback could have important consequences for the retention of gas and metals as well as for the chemical evolution and subsequent star formation in these systems.

The metallicity distributions of faint dwarf galaxies also provide important insights into their chemical evolution. Spectroscopic studies have revealed a broad range of stellar metallicities \citep{2007MNRAS.380..281M, 2007ApJ...670..313S}, including extremely metal-poor stars with $\mathrm{[Fe/H]} < -3$ \citep{Kirby_2008}. Several faint dwarf galaxies also show indications of multiple peaks in their metallicity distribution functions, suggesting the presence of chemically distinct stellar populations \citep{Brown_2014}. Different chemical evolution models are used to explain the different metallicity distribution functions \citep{2011ApJ...738...51L}. Together, these observations highlight the interplay between stellar feedback, gas retention and accretion, and metal enrichment in shaping the evolution of faint dwarf galaxies.

Observationally, the vast majority of extremely low-luminosity satellites detected are MW satellites. While a few have been discovered around other nearby massive galaxies \citep[e.g.,][]{2016ApJ...833..167M} and in the broader Local Volume as diffuse low surface brightness objects \citep[e.g.,][]{2017ApJ...850..109B, 2018ApJ...856...69D}, performing detailed stellar population analyses on such faint systems remains highly challenging. Examining their SFHs reveals a stark contrast with more massive dwarf galaxies, where star formation is typically quenched by environmental effects \citep{Geha_2012}. Even systems that were only recently accreted onto the MW halted their star formation billions of years ago \citep{fritz, Simon_2018}. This early quenching is widely attributed to the cosmic epoch of reionization \citep{Brown_2014, 2017MNRAS.471.3547F, Rodriguez_Wimberly_2018}, implying that star formation in these extremely low-mass systems was suppressed globally, regardless of their proximity to a massive host galaxy. An exception to this general picture of early quenching is the nearby satellite galaxy Leo T, which exhibits clear signatures of intermediate and even recent star formation.

Leo T is a low-mass, gas-rich MW satellite galaxy, with a brightness of $\mathrm{M_{v}}=-8$. This galaxy is slightly brighter than ultra-faint dwarf galaxies \cite[UFDs; $\mathrm{M_{v}}>-7.7$;][]{2019ARA&A..57..375S}. It is considered a transitional galaxy, with properties of both dwarf spheroidal (dSph) and dwarf irregular (dIrr) galaxies \citep{Irwin_2007}. Leo T is at a distance of 407$^{+29}_{-27}\,\mathrm{kpc}$ \citep{2012ApJ...756..108C}, and has a stellar mass of $2 \times 10^5\,\mathrm{M_{\odot}}$ \citep{deJong_2008}. Deep H\,\textsc{i} observations have revealed that the galaxy is gas-dominated, with a gas mass that is twice that of the stellar mass \cite[$\mathrm{M_{HI}=}4 \times 10^5\,\mathrm{M_{\odot}}$;][]{2018A&A...612A..26A}. Therefore, Leo T is one of the faintest star-forming, gas-rich galaxies. Such features are more commonly associated with dIrr galaxies \citep{10.1093/mnras/stae907, Hunter_2012}.

Several studies have explored the SFH of Leo T \citep{Irwin_2007, deJong_2008, Weisz_2012, 2012ApJ...756..108C, vaz2023}. In particular, optical photometry from HST \citep{Weisz_2012, 2012ApJ...756..108C} has revealed multiple distinct episodes of star formation within the galaxy. While 50\% of the stellar population formed more than 7 Gyr ago, a younger population with an age of less than 500 Myr is also present. Star formation activity has also been detected at early epochs (between 10 and 13 Gyr ago), although with lower star formation rates (SFRs) compared to later episodes.

Detailed H\,\textsc{i} studies have characterized the neutral gas in Leo T by decomposing it into cold neutral medium (CNM) and warm neutral medium (WNM), providing kinematic estimates for both phases \citep{10.1111/j.1365-2966.2007.12734.x, 2018A&A...612A..26A}. 
This multiphase structure is further supported by recent spectroscopic analysis using the Multi Unit Spectroscopic Explorer (MUSE). Specifically, \citet{vaz2023} find that the youngest stellar population exhibits a low velocity dispersion
($\sigma=2.31^{+2.68}_{-1.65}\,\mathrm{km s^{-1}}$), which is in excellent agreement with the kinematics of the CNM. This consistency suggests a direct physical link between the cold gas reservoir and the galaxy's recent star formation activity. 
The binary fraction is also estimated to be around 55$\%$ in another study \citep{vaz2026musefaintsurveyvbinary}. A recent study by \cite{2025A&A...699L..11V} reports the presence of molecular hydrogen near the center of Leo T with $1.4 \pm 0.4 \times 10^4\,\mathrm{M}_{\odot}$, which is 3\% of the total gas mass of this galaxy. 

The inner region of this galaxy exhibits minimal H$\alpha$ emission \citep{vaz2023} and some ultraviolet (UV) emission \citep{2015ApJ...808..136D}, both of which usually originate from massive stars. However, MUSE observations confirm that the H$\alpha$-emitting stars are associated with fast-rotating Be stars rather than massive stars, such as O or early B-type stars \citep{vaz2023}. Consequently, the star formation mechanisms operating in Leo T must differ significantly from those in typical disk galaxies. Given its exceptionally low stellar surface density and correspondingly weak self-gravity, understanding the origin of the observed UV emission in such an extreme, faint system presents a compelling astrophysical puzzle.

In this study, we utilized Ultraviolet Imaging Telescope (UVIT) far-UV (FUV) observations to extract UV star-forming clusters (SFCs) and characterize them. We also utilized the MUSE integral field unit (IFU) and catalog from \cite{vaz2023} and the publicly available HST photometric catalog from \cite{Weisz_2012} to gain insights into the FUV-emitting population in such extreme environments. We then compared our findings with those of previous studies.

In the following sections, we first present the observations used and data reduction in Sect. \ref{sec:observations}, then discuss the extracted FUV SFCs in Sect. \ref{UV Star forming Clusters}. In Sect. \ref{Possible UV-emitting stars}, we present the possible FUV-emitting stars for each SFC and discuss their distribution. In Sects. \ref{Radial trend} and \ref{interaction effect}, we discuss the radial variation of emission from young and evolved stars and also the effect of Leo T infall on star formation, respectively. We present Leo T's SFR and star formation efficiency (SFE) in Sect. \ref{SFR and SFE}.  Finally, we summarize and conclude our findings in Sect. \ref{sec Conclusions}.

%--------------------------------------------------------------------
\section{Observations}\label{sec:observations}

Leo T is part of the MUSE Faint survey \citep{Muse_faint}, a GTO survey of UFDs (PI Brinchmann). MUSE is an IFU installed at the Very Large Telescope (VLT), which provides a field of view (FoV) of 1 \(\times\) 1\,arcmin$^{2}$. This instrument has a spatial sampling of 0.2$^{\prime\prime}$ per pixel and a wavelength sampling of 1.25\,\r{A} per spectral channel. The details of the MUSE observation and reduction of this galaxy are described in \cite{vaz2023}.

We observed Leo T in the FUV band using the UVIT on board ASTROSAT \citep{2012SPIE.8443E..1NK}. Twin telescopes in UVIT can simultaneously observe in three bands: FUV (1300-1800\,\r{A}), near-ultraviolet (NUV; 2000-3000\,\r{A}), and the visible band (VIS; 3200-5500\,\r{A}). The spatial resolution of UVIT is around 1.2-1.5$^{\prime \prime}$ (with physical resolution of 2.37 to 2.97 pc). It has a FoV of 28$^\prime$ in diameter. Leo T was observed in filter F148W ($\mathrm{\lambda_{mean}}$=1481\,\r{A}, $\Delta\lambda$=500\,\r{A}). The total exposure time for these observations is 13.8 ks. We downloaded the raw level-1 data from the Indian Space Science Data Centre (ISSDC) and reduced it using the graphical user interface, CCDLAB \citep{2017PASP..129k5002P,2021JApA...42...30P}. 
We also used archival FUV and NUV data from the Galaxy Evolution Explorer \cite[GALEX;][]{2005ApJ...619L...1M}, which have spatial resolutions of 4.2 and 5.3$^{\prime \prime}$ (corresponding to 8.14 pc and 10.5 pc), respectively.

We used an archival Dark Energy Camera Legacy Survey (DECaLS) r-band image to determine whether the origin of UV emission is mainly from a younger or an older population. The Dark Energy Camera \cite[DECam;][]{2015AJ....150..150F} is mounted on the Blanco 4-meter telescope, located at the Cerro Tololo Inter-American Observatory (CTIO). It has a FoV of 3.2 square degrees and a median resolution of around 1.2 $^{\prime \prime}$.

\begin{figure*}
\sidecaption
  \includegraphics[width=12cm]{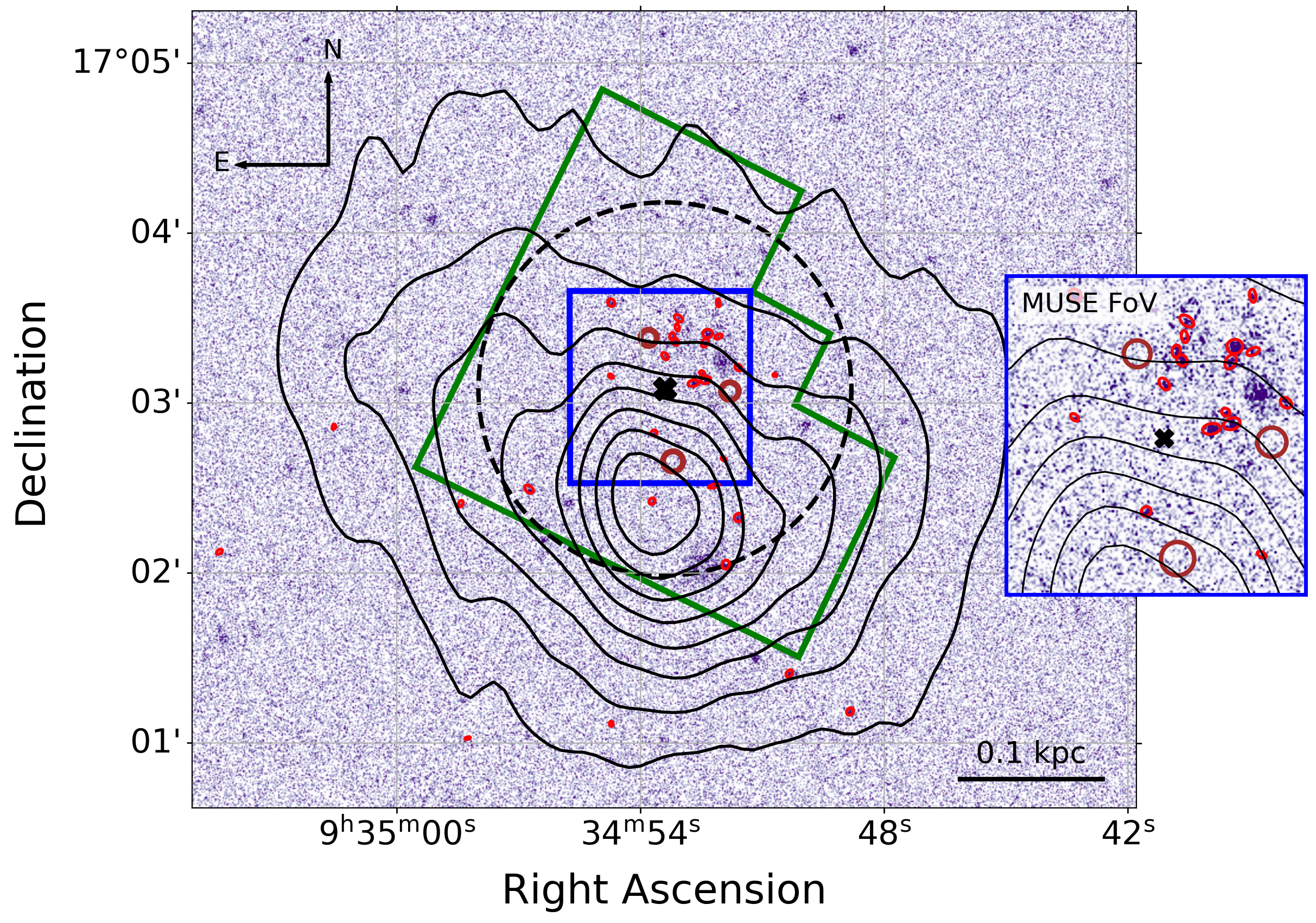}
     \caption{Background: UVIT FUV image of Leo T. The optical center is marked as a black cross. The red ellipses are the extracted FUV SFCs. The black contours represent H\,\textsc{i} intensity contours from deeper Westerbork Synthesis Radio Telescope observations \citep{2018A&A...612A..26A}, ranging from 20\% to 90\% of the peak column density with a step size of 10\%. The peak H\,\textsc{i} column density is 4.2 $\times$10$^ {20}$cm${^{-2}}$. The green polygon represents the HST FoV \citep{Weisz_2012}, and the blue polygon represents the MUSE FoV \citep{vaz2023}. The dashed black circle represents the FoV of the Atacama Compact Array, and the brown circles are the molecular cloud locations, as mentioned in \cite{2025A&A...699L..11V}. The MUSE FoV is shown in greater detail in the zoomed-in panel.}
     \label{fig:leo_patches}
\end{figure*}

Distribution of UV clusters can be seen in Fig. \ref{fig:leo_patches}. This figure also compares the HST and MUSE FoVs with respect to the inner H\,\textsc{i} intensity contours and UV extent of Leo T. The plotted H\,\textsc{i} intensity contours are from deep Westerbork Synthesis Radio Telescope observations \citep{2018A&A...612A..26A}.

\section{FUV star-forming clusters}\label{UV Star forming Clusters}

 \begin{figure}
   \centering
   \includegraphics[scale=0.35]{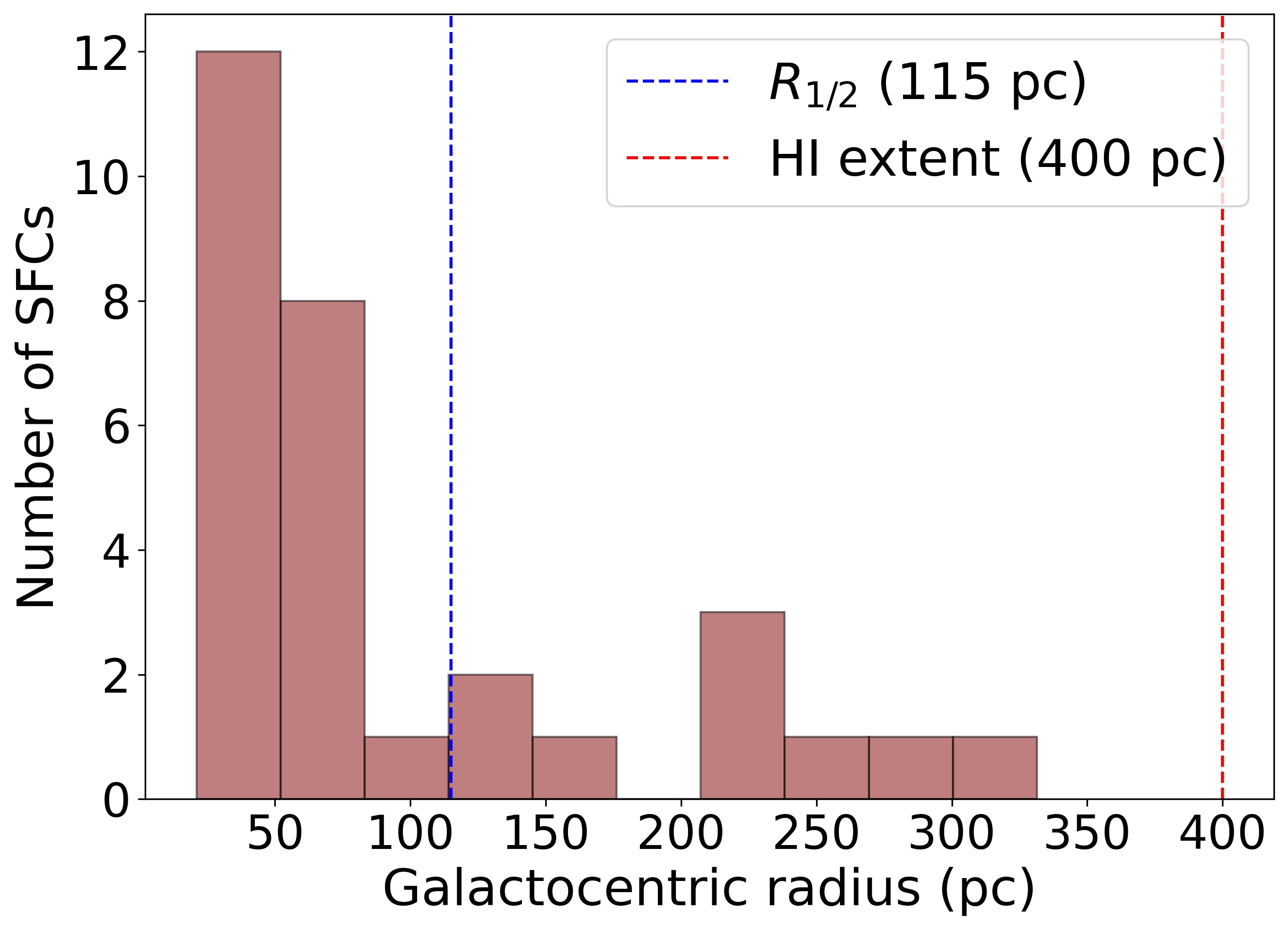}
   \caption{Distribution of SFCs along the galactocentric radius. The vertical blue and red lines represent the half light radius R$_{1/2}$=115 pc \citep{10.1111/j.1365-2966.2010.16753.x} and H\,\textsc{i} extent of the galaxy \cite[400pc;][]{2018A&A...612A..26A}, respectively. }
              \label{fig:SFCs_extent}
    \end{figure}

We performed background subtraction of the science-ready UVIT FUV image using IRAF \citep{1993ASPC...52..173T}, as mentioned in \cite{10.1093/mnras/stae907}. We used Source Extractor \cite[SExtractor;][]{1996A&AS..117..393B} to extract SFCs, adopting a detection threshold of $2\,\sigma$, where $\sigma$ represents the estimated global background noise of the FUV image. For Leo T FUV image, $\sigma$ is about 0.454 counts. A relatively low threshold was chosen because no significant diffuse FUV emission is associated with Leo T. We used a minimum detection area of 10 pixels. Standard deblending threshold and deblending constant of 32 and 0.001 were adopted based on visual inspection. By running these configurations, we obtain the positions and ellipse parameters of extracted sources.

Foreground sources were masked by cross-matching the extracted sources with TOPCAT \citep{2005ASPC..347...29T} using Gaia Data Release 3 (DR3) data \citep{Gaia, refId0}. We also masked background extended sources by comparing the UVIT field with the optical Sloan Digital Sky Survey DR18 field (gri color composite) \citep{Almeida_2023}. In total, we identified 30 SFCs in Leo T, which are marked by red ellipses in Fig.~\ref{fig:leo_patches}. 

\begin{figure}
   \centering
   \includegraphics[scale=0.39]{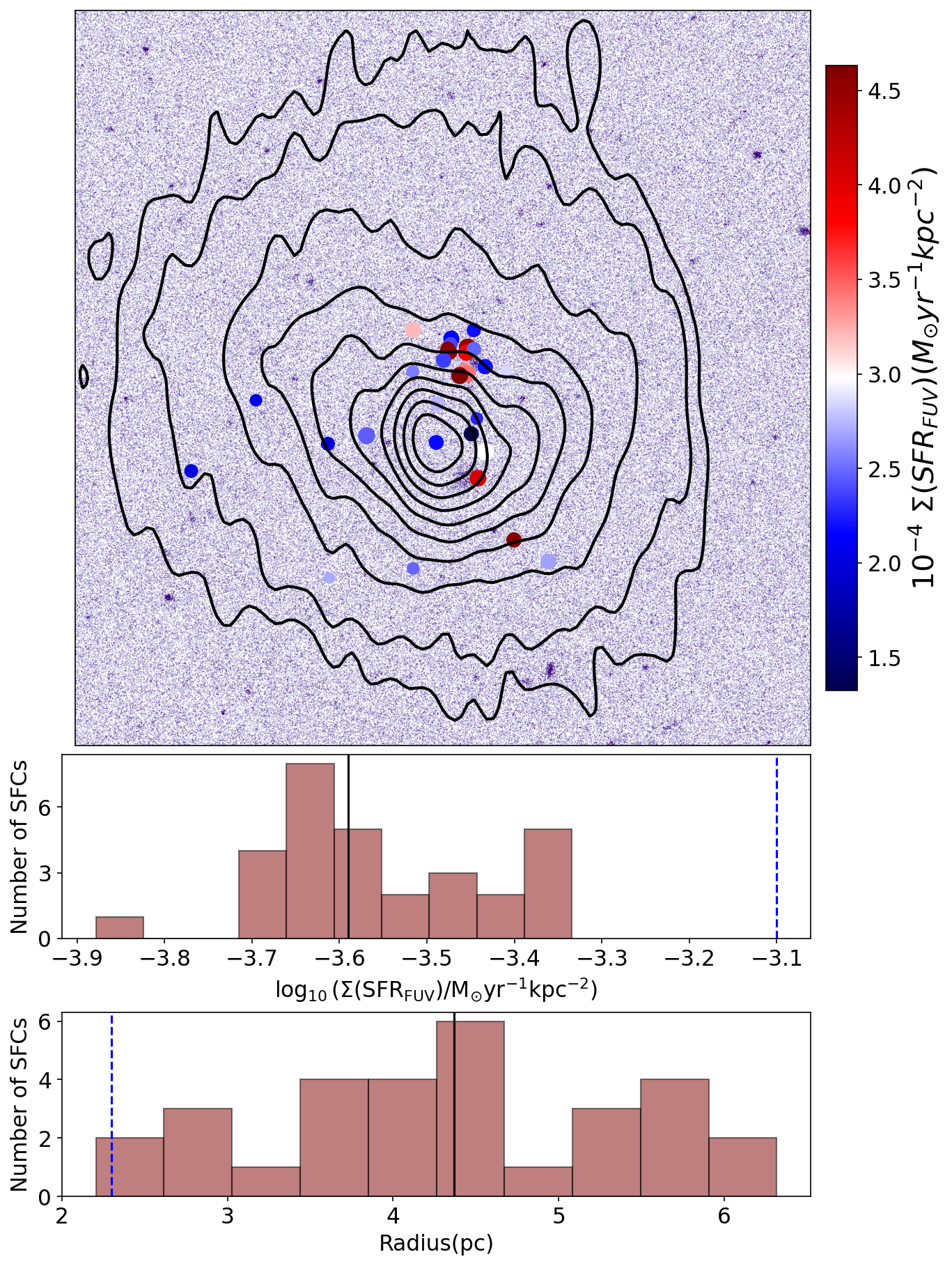}

   \caption{Background: UVIT FUV image of Leo T. The black contours represents the H\,\textsc{i} intensity contours \citep{2018A&A...612A..26A}, ranging from 5\%, 10\% to 90\% of the peak column density with a step size of 10\%. The peak H\,\textsc{i} column density is 4.2 $\times$10$^ {20}$cm${^{-2}}$. The image shows how $\Sigma_{\rm SFR}$ and the radius of SFCs are distributed in the galaxy. The image also includes radius and $\Sigma_{\mathrm{SFR}}$ histograms. The solid black lines represent the median values. The dashed line in the radius histogram is the resolution limit and is the lower $\Sigma _{\mathrm{SFR}}$ limit \citep{10.1093/mnras/stae907} for dIrrs in the $\Sigma_{\mathrm{SFR}}$ histogram.}
              \label{fig:leot_properties}%
    \end{figure}

\subsection{Extent of FUV SFCs in Leo T}\label{Extent of UV SFCs}

We first compared the distribution of FUV-emitting SFCs with respect to the H\,\textsc{i} extent. In Fig. \ref{fig:leo_patches}, we can clearly see that most of the SFCs are clustered near the optical center (RA = 143.722500 deg, Dec = 17.051389 deg), with no SFCs detected farther north. About 70\% of the SFCs lie within the half-light radius of Leo T, $R_{1/2}=115\pm17$\,pc, derived by \citet{10.1111/j.1365-2966.2010.16753.x} from the elliptical surface-brightness fit reported by \citet{deJong_2008}, based on optical $g$- and $r$-band observations with the Large Binocular Telescope.
 The H\,\textsc{i} extent of the galaxy is approximately 400\,pc, which is nearly four times R$_{1/2}$ \citep{2018A&A...612A..26A}. While the central regions contain most of the identified SFCs, those located at larger galactocentric radii are preferentially distributed toward the south and east of both the optical and H\,\textsc{i} centers. Figure \ref{fig:SFCs_extent} presents the distribution of SFCs as a function of galactocentric radius, whereas Fig. \ref{fig:leo_patches} illustrates their asymmetric spatial distribution.

\subsection{Properties of the FUV SFCs}\label{Properties of UV Star forming Clusters}

We can resolve star clusters above a size of 2.37\,pc (dashed blue line in Fig. \ref{fig:leot_properties}), which corresponds to the UVIT resolution of 1.2$^{\prime\prime}$ at the distance of 409\,kpc. The radii of the SFCs in Leo T are above this limit. This physical extent is comparable to the typical scale of bound star clusters \citep{2014ApJ...787L..15E}. Consequently, these SFCs may host multiple FUV-emitting stars. We see in Fig. \ref{fig:leot_properties} that the median radius of the SFCs is approximately 4.4\,pc (solid black line). There is not much difference in size as we go from the inner region to the outer region of the galaxy.

\begin{figure}
   \centering
   \includegraphics[scale=0.33]{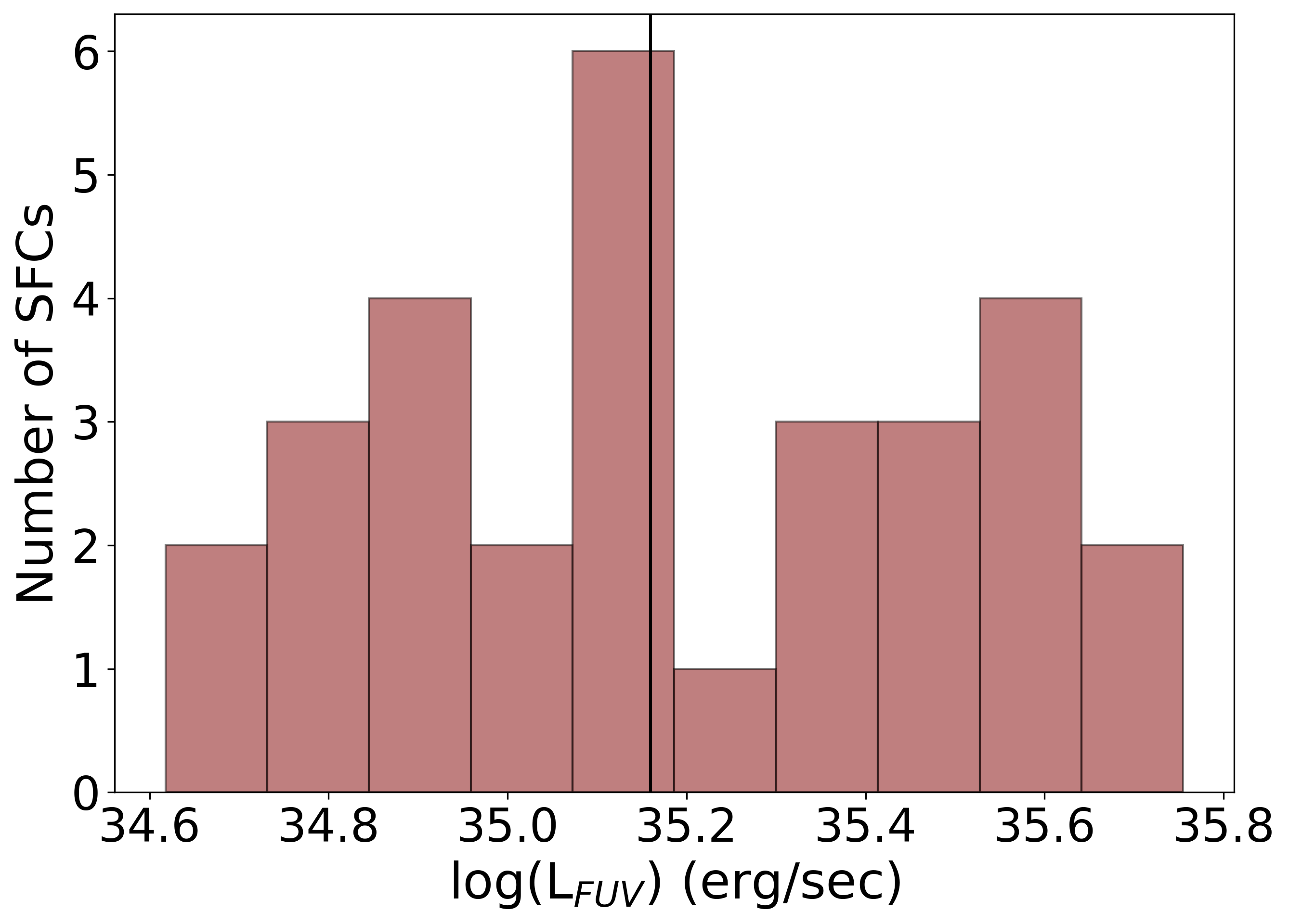}
   \caption{Histogram of the FUV luminosity of SFCs. The solid black line represents the median value.}
    \label{fig:leot_luminosity}
    \end{figure}

 We determined the total counts within the SFCs and corrected for MW extinction for the FUV image. The Galactic reddening, $\mathrm{E(B-V)}$, was obtained from Infrared Science Archive (IRSA) \citep{2011ApJ...737..103S}. The corresponding FUV extinction, $A_{\rm FUV}$, was estimated using the filter-specific ratio $A_{\rm FUV}/A_V$ provided by \cite{2020sea..confE.182R}. Given the negligible dust content of the galaxy \citep{2015ApJ...808..136D}, we did not apply corrections for internal extinction. We converted counts per second to flux using the conversion factor from \cite{Tandon_2020}. The FUV luminosity of these SFCs is distributed between $10^{34}\,\mathrm{ergs^{-1}}$ to $10^{36}\,\mathrm{ergs^{-1}}$. Using the \citet{castelli2004newgridsatlas9model} stellar atmosphere models, we see that the estimated FUV luminosity of a single massive B star exceeds $10^{36}\,\mathrm{ergs^{-1}}$ \citep{Amrutha_2025}. In contrast, the median FUV luminosity of the SFCs in Leo T is only $\sim 10^{35.16}\,\mathrm{ergs^{-1}}$ (the full distribution is shown in Fig. \ref{fig:leot_luminosity}). This clear discrepancy allows us to confidently rule out the presence of individual massive OB stars in these complexes.

We calculated the FUV SFR for the SFCs using 
\begin{equation}
\mathrm{SFR}_{\mathrm{FUV}}=0.68 \times 10^{-28} \times \mathrm{L_{FUV}}
\label{eq1}
\end{equation}
 where \(\mathrm{L_{FUV}}\) is the extinction corrected FUV luminosity in \(\mathrm{ergs^{-1}Hz^{-1}}\). This relation is from \cite{Salim_2007} for subsolar metallicity and Chabrier IMF \citep{Chabrier_2003}. 
Far-ultraviolet-derived SFR traces star formation over a timescale of $\sim100$ Myr, whereas H$\alpha$ traces star formation over only the past $\sim10$ Myr \citep{ 2004A&A...421..887I}. Consequently, FUV emission is sensitive to a larger stellar population and longer-term star formation activity than H$\alpha$. Given the very low metallicity of Leo T, the FUV-based SFR may be slightly overestimated. Furthermore, the standard FUV-SFR calibration assumes a fully sampled IMF and a constant recent SFH, which may not hold in a low-mass system such as Leo T. Therefore, the derived SFRs should be regarded as approximate and subject to additional systematic uncertainties.

 The estimated FUV SFR density $\Sigma (\rm SFR_{FUV})$ of SFCs lies between 10$^{-3.9}\,\mathrm{M_{\odot}yr^{-1}kpc^{-2}}$ to 10$^{-3.3}\,\mathrm{M_{\odot}yr^{-1}kpc^{-2}}$. We compared these values with the $\Sigma (\rm SFR_{FUV})$ of SFCs in star-forming dwarf galaxies (SFDGs), primarily dIrrs. The study by \cite{10.1093/mnras/stae907} showed that SFCs of dIrrs exhibit substantially lower $\Sigma (\rm SFR_{FUV})$ than other SFDG types, with 10$^{-3.09}\,\mathrm{M_{\odot}yr^{-1}kpc^{-2}}$ being the lowest value reported in their sample. Even the highest $\Sigma (\rm SFR_{FUV})$ measured for the Leo T SFCs falls below this value, highlighting the exceptionally inefficient star formation occurring in this low-mass satellite galaxy. Such low values are consistent with the low stellar surface density of Leo T compared with those of SFDGs \citep{deJong_2008, 10.1093/mnras/stae907}.

 \section{Possible FUV-emitting stars in Leo T}\label{Possible UV-emitting stars}

\subsection{Cross-matching with HST catalog}\label{Cross-matching with HST catalog}
To investigate the stellar content of the SFCs, we utilized the HST photometric catalog of well-measured stars published by \citet{Weisz_2012}. The catalog provides precise spatial coordinates of stars, as well as high-fidelity photometry (F814W and F606W magnitudes, including associated errors). We then performed a spatial cross-match between our extracted SFCs and the HST stellar sources. The catalog of FUV clusters, associated HST stars, and their ages are given in Table \ref{Table:UV_HST_MUSE}. Of the 30 identified UV SFCs, 22 fall within the HST FoV. However, we identified potential FUV-emitting stellar counterparts for only 19 of these SFCs. The remaining three (UV sources 7, 9, and 14 in Table \ref{Table:UV_HST_MUSE}) are located in under-dense regions of the HST footprint and do not spatially coincide with any resolved stars capable of producing the observed FUV emission. This under-density is found in the regions where two HST fields are combined to form a mosaic image. 

\begin{figure}
   \centering
   \includegraphics[scale=0.30]{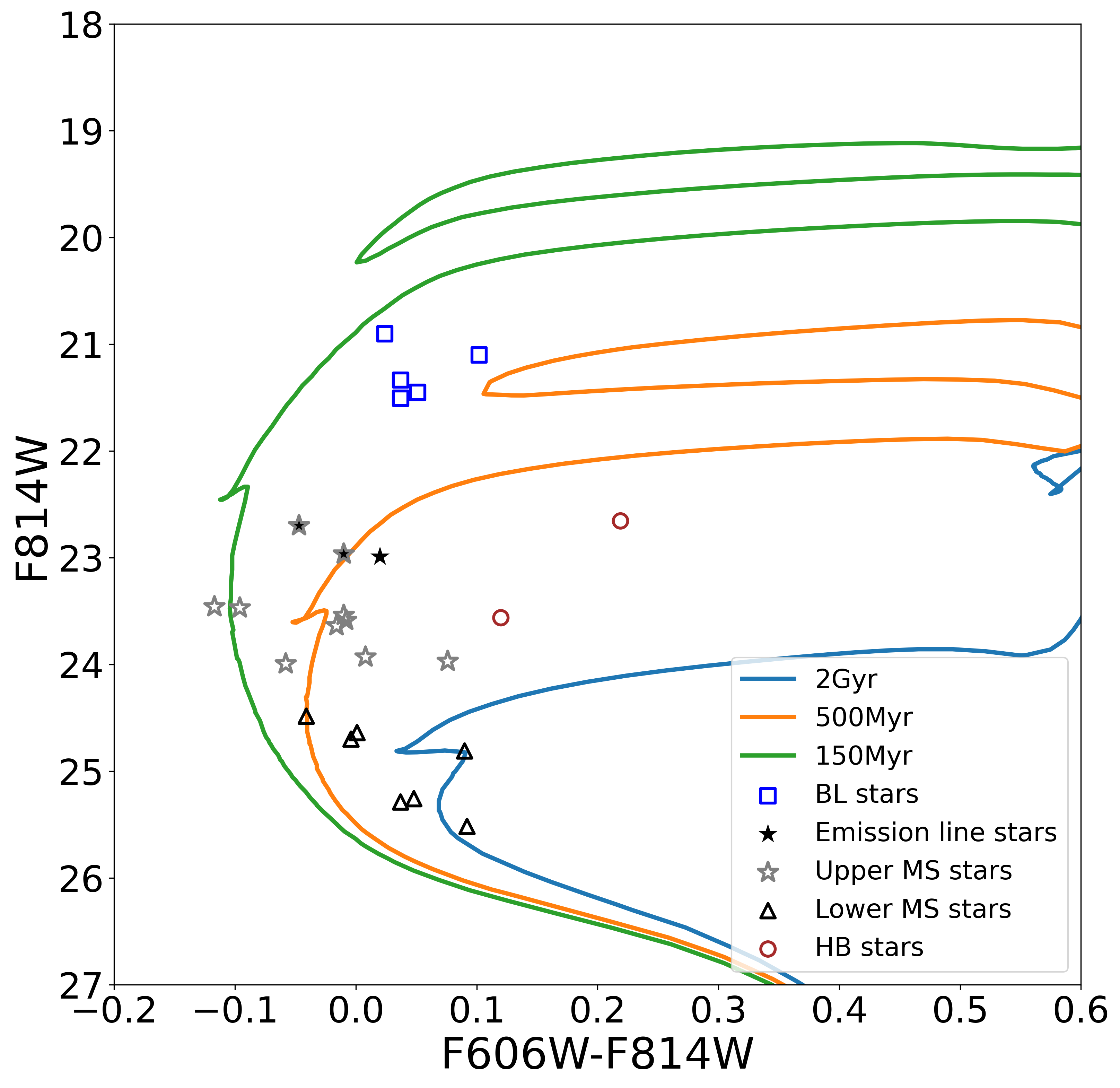}

   \caption{Deep HST/WFPC2 CMD of Leo T with select Padova isochrones for FUV sources. Isochrones of different colors represent different ages, and markers of different colors represent different populations. }
              \label{fig:cmd_final}%
    \end{figure}

We plotted all the associated HST stars in a color-magnitude diagram (CMD) to identify the possible FUV-emitting stars. We fitted isochrones of different ages from 2 Gyr to 150 Myr. We used Padova's stellar evolution models \citep{2012MNRAS.427..127B} and adopted the parameters from \cite{Weisz_2012}. We adopted a fixed metallicity of $\mathrm{[M/H]} = -1.6$, as \citet{vaz2023} demonstrated that the stellar populations in Leo T are consistent with this value regardless of their age. We removed all the stars with F814W $>$ 25.5 and F606W-F814W $>$ 0.5, as these are the faintest and reddest stars, and least likely to emit FUV. The CMD with possible FUV-emitting HST stars is shown in Fig. \ref{fig:cmd_final}.

To classify the resolved HST counterparts, we compared their photometry with the frameworks established by \citet{Weisz_2012} and \citet{2012ApJ...756..108C}. We find that these FUV-emitting complexes are primarily populated by blue loop (BL) stars, upper and lower main-sequence (MS) stars, and a few horizontal branch (HB) stars. Notably, 79\% (15 out of 19) of these SFCs contain only a single resolved FUV-emitting star. Breaking down the stellar types, five SFCs host BL stars, while nine harbor upper MS stars (including three emission-line stars, detailed in the following section). Furthermore, lower MS stars are present in seven SFCs, three of which are co-located with upper MS companions. In terms of stellar ages, 95\% of the SFCs are younger than 1\,Gyr. Within this young population, 11 SFCs host stars younger than 500\,Myr, and only two contain stars younger than 200\,Myr. Conversely, the SFCs located near the H\,\textsc{i} center host two older HB stars with estimated ages of $\sim 2$\,Gyr.

\subsection{Cross-matching with MUSE sources}\label{Cross-matching with MUSE sources}
\cite{vaz2023} identified 58 stars as members of Leo T using MUSE IFU data, classifying 20 stars as young, with age $\leq$1 Gyr. We cross-matched the FUV-emitting SFCs with these 58 stars and also with the HST catalog. Only six stars have FUV emission. Five are classified as upper MS stars when compared with the HST catalog stars. Three of these are H$\alpha$-emitting stars, also known as emission-line (EM) stars. These three stars are identified as Be stars by \cite{vaz2023}. There is also one BL star in the MUSE catalog. We find that all the FUV-emitting MUSE stars have ages less than 500 Myr, and one star is younger than 200 Myr. 

To characterize the FUV-emitting stars identified in the MUSE catalog by \citet{vaz2023}, we derived their effective temperature, metal abundance, and line-of-sight velocity. These parameters were extracted using Spexxy, a full-spectrum fitting code \citep{2016A&A...588A.148H} that interpolates over a stellar spectral model grid. An analysis of these results reveals that the MUSE counterparts within the SFCs are uniformly hot, with temperatures exceeding $7000\,\text{K}$. Interestingly, among the 11 MUSE sources that meet this temperature threshold, only six exhibit FUV emission. To understand this dichotomy, we investigated their remaining parameters. We find that five out of the six FUV emitters are extremely metal-poor ([Fe/H]$\leq$-3). The single exception is an upper MS star (UV source 13), which displays a higher metallicity of $\sim -1.5$ and a substantially low line-of-sight velocity ($v_{\text{los}} = -11.84 \pm 27.99\,\text{km s}^{-1}$), in stark contrast to the $25$ to $50 \, \text{km s}^{-1}$ range observed for the other five stars. The velocity of this source also differs substantially from the mean stellar velocity and the velocity of the bulk of the H\,\textsc{i} emission, which peaks at $\sim39\, \mathrm{km\,s^{-1}}$ \citep{vaz2023, 2018A&A...612A..26A}. We note, however, that the generally low signal-to-noise ratio (S/N $\sim$ 3.47) of the MUSE spectra for this faint target might contribute to this kinematic and chemical discrepancy.

The presence of extremely metal-poor young stars is also qualitatively consistent with the chemical-evolution analysis of \citet{2011ApJ...738...51L}. This model marginally favors an extra gas scenario, in which an additional reservoir of pristine gas becomes available during the star-formation history, through either external infall or an internal source of cooling gas. However, the metallicities of the young stars are subject to large spectroscopic uncertainties, which preclude establishing a statistically significant difference between the young and old populations \citep{vaz2023}.

Recent work by \citet{vaz2026musefaintsurveyvbinary} indicates that the younger stellar population in the MUSE sample exhibits a significantly higher binary fraction (35\%) compared to the older population (15\%). Given that all FUV-emitting MUSE stars are young, they inherently possess a higher probability of residing in binary systems. However, cross-referencing our sources with the binary catalog from \citet{vaz2026musefaintsurveyvbinary} reveals only a single match: a BL star, and even for this target, the computed binary likelihood is low. Furthermore, identifying binaries among BL stars requires caution; these stars frequently exhibit radial and non-radial pulsations \citep{1979MNRAS.189..649B, Saio_2013}. Such intrinsic pulsations can induce subtle to significant variations in radial velocity, potentially mimicking the kinematic signatures of a binary companion.

\begin{table*}[]
\centering
\caption{Properties of FUV star-forming complexes}
\label{Table:UV_HST_MUSE}
\resizebox{\textwidth}{!}{%
\begin{tabular}{ccccccccccccccc}
\hline\\
UV & UV RA       & UV Dec      & Area               & Luminosity              & $\Sigma (\rm SFR_{FUV})$                      & HST RA      & HST Dec     & Color  & F814W  & Age     & MUSE RA     & MUSE Dec    & Star   \\

 SFC     & [deg]       & [deg]       & [10$^{-5}$kpc$^2$] & [10$^{35}$erg s$^{-1}$] & [10$^{-3}$ & [deg]       & [deg]       &        & [mag]  &  [Gyr]       & [deg]       & [deg]       & Type   \\
      &     &  &  &  & M$_{\odot}$yr$^{-1}$kpc$^{-2}$] &   &      &        &  &         &     & &   \\

\hline\\
\hline\\

1     & 143.7170179 & 17.05982212 & 4.66               & 1.03 $\pm$ 0.01         & 0.22  $\pm$ 0.003                         & 143.7167917 & 17.05961111 & -0.016 & 23.637 & $<$0.5 &             &             & Up MS  \\
2     & 143.7280098 & 17.05983809 & 8.17               & 2.63 $\pm$ 0.03         & 0.32 $\pm$ 0.003                          & 143.72775   & 17.05963889 & -0.01  & 22.965 & $<$0.5 & 143.7279075 & 17.05968746 & EM     \\
3     & 143.7210927 & 17.05832311 & 8.38               & 1.78 $\pm$0.02          & 0.21  $\pm$ 0.002                         & 143.72125   & 17.05883333 & -0.041 & 24.484 & $<$0.5 &             &             & Low MS \\
3     &             &             &                    &                         &                                           & 143.7207083 & 17.05805556 & -0.008 & 23.587 & $<$0.5 &             &             & Up MS  \\
4     & 143.7212312 & 17.05740724 & 4.39               & 1.05 $\pm$ 0.01         & 0.24  $\pm$ 0.003                         & 143.7210833 & 17.05727778 & 0.001  & 24.638 & $<$1   &             &             & Low MS \\
5     & 143.721749  & 17.05654516 & 5.59               & 2.58 $\pm$0.02          & 0.46 $\pm$ 0.005                          & 143.7214167 & 17.05633333 & 0.051  & 21.45  & $<$0.5 &             &             & BL     \\
6     & 143.7213978 & 17.05599662 & 6.3                & 2.85 $\pm$0.03          & 0.45  $\pm$ 0.004                         & 143.721125  & 17.05572222 & -0.047 & 22.7   & $<$0.5 & 143.7212683 & 17.05580793 & EM     \\
7     & 143.7181246 & 17.05682249 & 12.53              & 5.69 $\pm$ 0.05          & 0.45 $\pm$ 0.004                            &             &             &        &        &        &             &             &        \\
8     & 143.7183826 & 17.05591634 & 8.75               & 3.39 $\pm$0.03          & 0.38  $\pm$ 0.004                         & 143.718125  & 17.05577778 & 0.048  & 25.257 & $<$0.2 &             &             & Low MS \\
8     &             &             &                    &                         &                                           & 143.7179167 & 17.05658333 & -0.117 & 23.459 & $<$0.2 &             &             & Up MS  \\
8     &             &             &                    &                         &                                           & 143.7182083 & 17.05547222 & -0.058 & 23.994 & $<$0.5 &             &             & Up MS  \\
9     & 143.7169908 & 17.05652612 & 5.24               & 1.31 $\pm$0.01          & 0.25  $\pm$ 0.003                         &             &             &        &        &        &             &             &        \\
10    & 143.7224884 & 17.05460462 & 6.47               & 1.53 $\pm$0.02          & 0.23  $\pm$ 0.003                         & 143.722875  & 17.05427778 & 0.076  & 23.97  & $<$1   &             &             & Up MS  \\
11    & 143.714962  & 17.05349263 & 6.55               & 1.44 $\pm$0.02          & 0.22  $\pm$ 0.002                         & 143.7147917 & 17.05333333 & -0.01  & 23.539 & $<$0.5 & 143.7148875 & 17.05338229 & Up MS  \\
12    & 143.7112194 & 17.05275168 & 1.83               & 0.52 $\pm$0.01          & 0.28  $\pm$ 0.005                         & 143.7109167 & 17.05266667 & 0.008  & 23.93  & $<$1   &             &             & Up MS  \\
13    & 143.728043  & 17.05263533 & 3.02               & 0.78 $\pm$0.01          & 0.26  $\pm$ 0.003                         & 143.727875  & 17.05233333 & -0.004 & 24.699 & $<$1   &             &             & Low MS \\
13    &             &             &                    &                         &                                           & 143.7277917 & 17.0525     & -0.096 & 23.47  & $<$0.2 & 143.7280032 & 17.0524698  & Up MS  \\
14    & 143.7183057 & 17.05225203 & 10.21              & 3.53 $\pm$0.03          & 0.34  $\pm$ 0.003                         & 143.7180417 & 17.05219444 & 0.102  & 21.101 & $<$0.5 & 143.7181491 & 17.05218401 & BL     \\
15    & 143.7195777 & 17.05195244 & 10.5               & 4.9 $\pm$0.05           & 0.46  $\pm$ 0.004                         & 143.7202083 & 17.05186111 & 0.037  & 25.288 & $<$1   &             &             & Low MS \\
15    &             &             &                    &                         &                                           &             &             &        &        &        & 143.719469  & 17.05177861 & EM     \\
16    & 143.7236156 & 17.04707352 & 5.26               & 1.45 $\pm$0.02          & 0.27  $\pm$ 0.003                         & 143.7235    & 17.04752778 & 0.092  & 25.517 & $<$1   &             &             & Low MS \\
17    & 143.716462  & 17.04451533 & 2.8                & 0.64 $\pm$0.01          & 0.23  $\pm$ 0.003                         &             &             &        &        &        &             &             &        \\
18    & 143.7174775 & 17.04187339 & 6.14               & 0.82 $\pm$0.01          & 0.13  $\pm$ 0.002                         & 143.717375  & 17.04166667 & 0.09   & 24.81  & $<$1   &             &             & Low MS \\
19    & 143.7364467 & 17.04155411 & 10.41              & 2.57 $\pm$0.02          & 0.25  $\pm$ 0.002                         & 143.7361667 & 17.04138889 & 0.037  & 21.506 & $<$0.5 &             &             & BL     \\
20    & 143.7238222 & 17.04037914 & 5.85               & 1.29 $\pm$0.01          & 0.22  $\pm$ 0.002                         & 143.723625  & 17.04033333 & 0.12   & 23.562 & $<$2   &             &             & HB     \\
20    &             &             &                    &                         &                                           & 143.723875  & 17.04022222 & 0.219  & 22.655 & $<$2   &             &             & HB     \\
21    & 143.7149312 & 17.03878537 & 11.32              & 3.4 $\pm$0.03           & 0.3  $\pm$ 0.003                          & 143.7146667 & 17.03872222 & 0.024  & 20.901 & $<$0.5 &             &             & BL     \\
22    & 143.7162566 & 17.03417653 & 9.86               & 4.03 $\pm$0.04          & 0.41  $\pm$ 0.004                         & 143.7159583 & 17.03405556 & 0.037  & 21.336 & $<$0.5 &             &             & BL     \\
23    & 143.7280064 & 17.018557   & 2.79               & 0.69 $\pm$0.01          & 0.25  $\pm$ 0.004                         &             &             &        &        &        &             &             &        \\
24    & 143.7431595 & 17.01691507 & 1.53               & 0.42$\pm$0.01           & 0.27  $\pm$ 0.006                         &             &             &        &        &        &             &             &        \\
25    & 143.7097425 & 17.0235024  & 6.52               & 3 $\pm$0.03             & 0.46  $\pm$ 0.004                         &             &             &        &        &        &             &             &        \\
26    & 143.7035042 & 17.01977132 & 7.68               & 2.07 $\pm$0.02          & 0.27  $\pm$ 0.003                         &             &             &        &        &        &             &             &        \\
27    & 143.7682088 & 17.03541104 & 4.12               & 0.83 $\pm$0.01          & 0.20  $\pm$ 0.003                          &             &             &        &        &        &             &             &        \\
28    & 143.7434459 & 17.04014448 & 4.37               & 0.86 $\pm$0.01          & 0.20  $\pm$ 0.003                          &             &             &        &        &        &             &             &        \\
29    & 143.7564737 & 17.04766672 & 2.76               & 0.56 $\pm$0.01          & 0.20  $\pm$ 0.003                         &             &             &        &        &        &             &             &        \\
30    & 143.7186868 & 17.05293903 & 4.15               & 1.41 $\pm$0.01          & 0.34  $\pm$ 0.004                        &             &             &        &        &        &             &             &       \\

\hline
\end{tabular}%
}
\vspace{0.4cm}

\begin{minipage}{180mm}
Note: UV RA, UV Dec, and area are SFC parameters obtained using SExtractor from UVIT FUV observation. $\Sigma (\rm SFR_{FUV})$ is estimated FUV SFR density as mentioned in Sect. \ref{Properties of UV Star forming Clusters}. HST RA, HST Dec, color (F606W-F814W), and magnitude F814W are the stellar parameters from the HST catalog \cite{Weisz_2012} for cross-matched stars. Age denotes the age of HST stars. Age and star types are inferred from the CMD (See Fig. \ref{fig:cmd_final}). MUSE RA and MUSE Dec are stellar parameters of cross-matched stars from \cite{vaz2023}.
\end{minipage}
\end{table*}

\subsection{Distribution of FUV-emitting stars in Leo T}\label{Distribution of UV-emitting stars}
\begin{figure}
   \centering
   \includegraphics[scale=0.30]{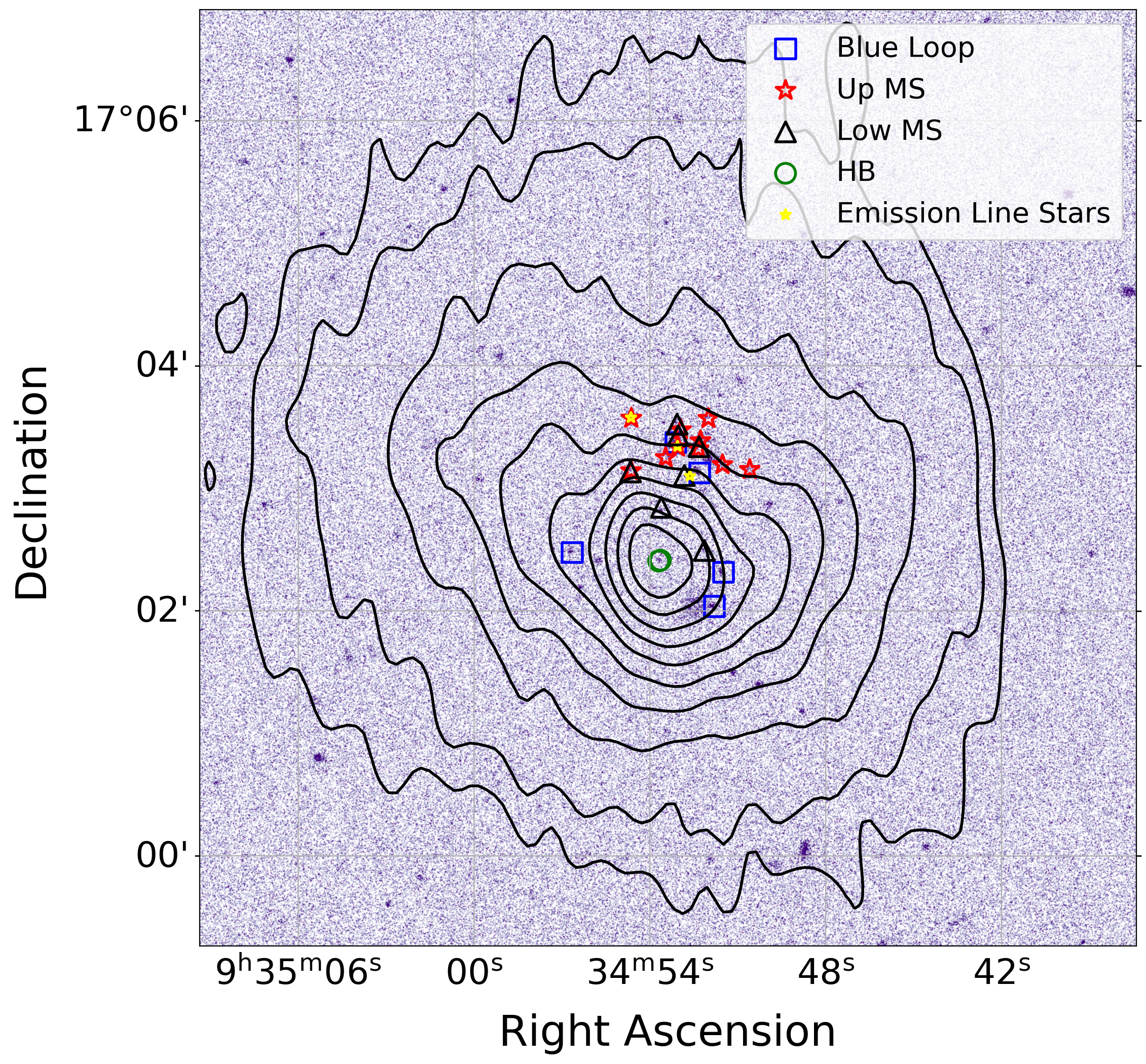}
   \caption{Distribution of the FUV-emitting stellar population of Leo T. Background: UVIT FUV image of Leo T. The black contours are H\,\textsc{i} contours \citep{2018A&A...612A..26A}. Different markers in the image represent different stellar populations emitting FUV.}
              \label{fig:leot_stars}%
    \end{figure}

We can infer the propagation of star formation in Leo T by examining the distribution of stellar populations of different ages. We analyzed the spatial distribution of the identified stellar classes relative to the optical and H\,\textsc{i} centers of Leo T. As illustrated in Fig. \ref{fig:SFCs_extent}, the majority of the SFCs are concentrated near the optical center. Breaking down this distribution by stellar type reveals a clear spatial segregation: all upper MS and EM stars are strictly confined to this central optical region, alongside a single BL star and several lower MS stars. In contrast, further south near the H\,\textsc{i} kinematic center, the stellar population shifts, becoming populated primarily by HB, BL, and additional lower MS stars. This distinct morphological gradient (further detailed in Fig. \ref{fig:leot_stars}) strongly indicates that the most recent episodes of star formation in Leo T were highly localized around its optical core.

\begin{figure}
   \centering
   \includegraphics[scale=0.31]{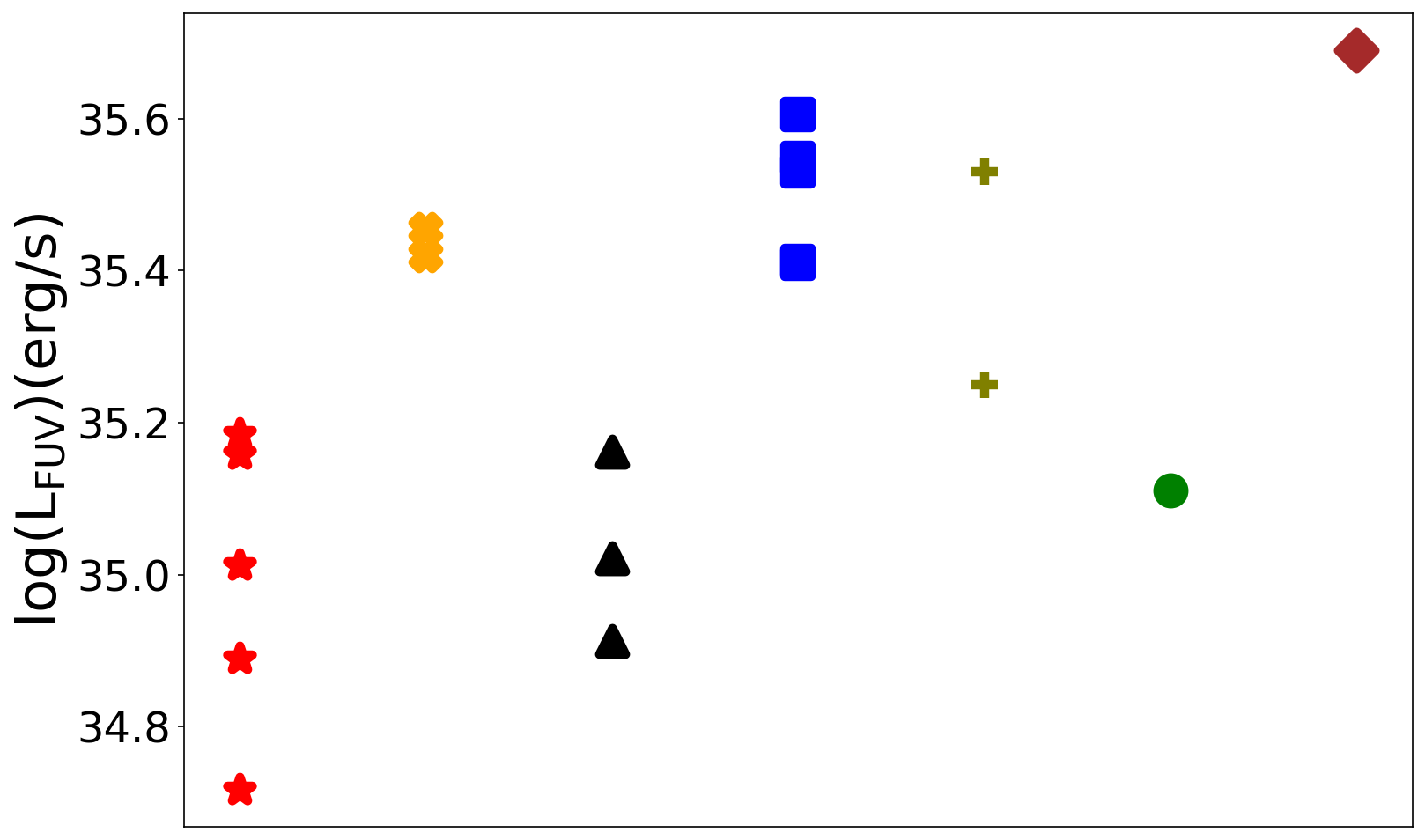}
   \includegraphics[scale=0.305]{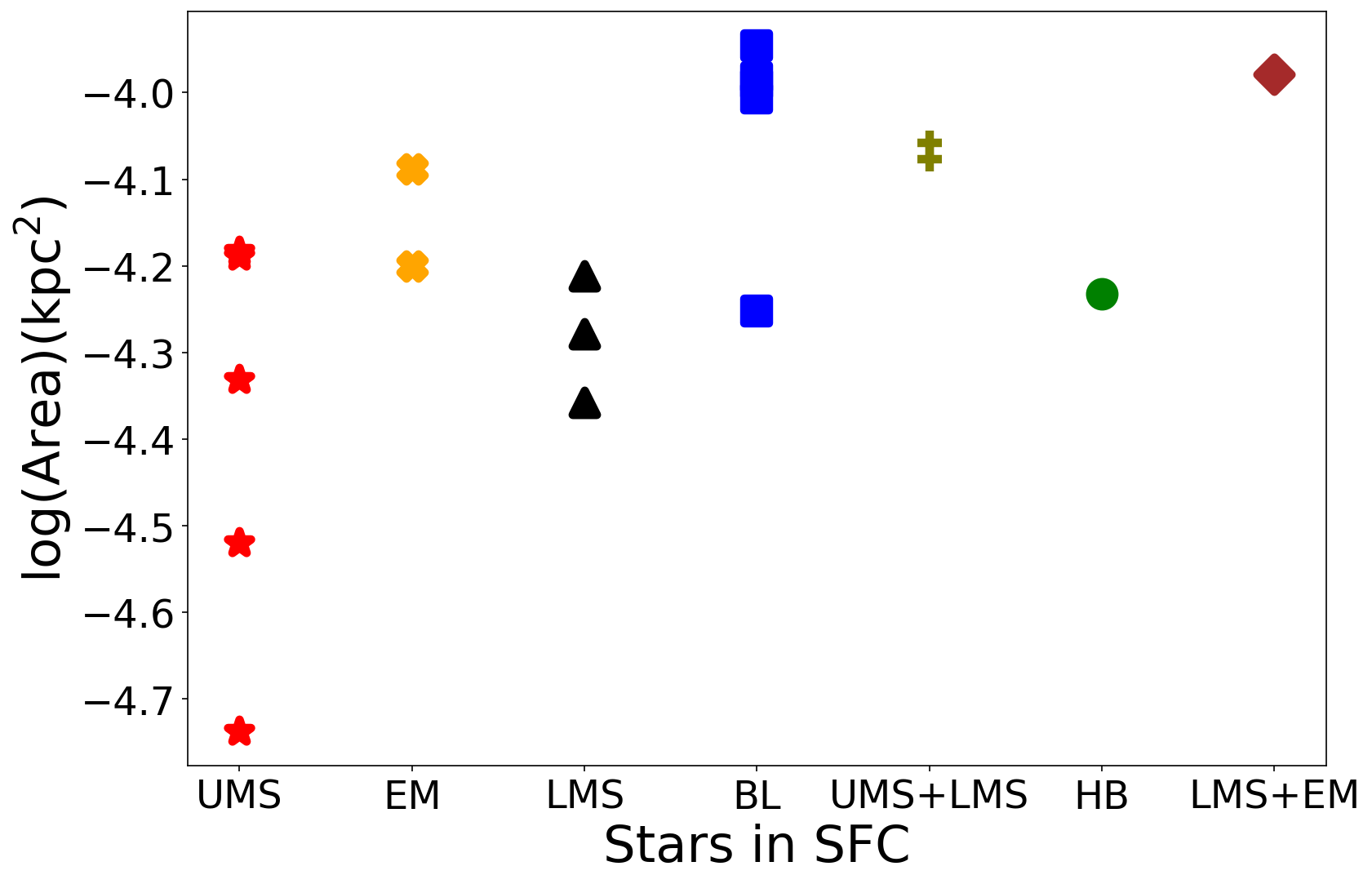}
   \caption{FUV luminosity (Top) and area distribution of SFCs relative to the type of star present in it. The stars and triangles represent SFCs with upper MS and lower MS stars, respectively. The plus represents
SFCs containing both lower and upper MS stars. BL and HB are represented by the square and circular markers, respectively. The SFCs with crosses and diamonds contain EM stars and both EM and lower MS stars, respectively}
              \label{fig:leot_stars_prop}%
    \end{figure}

Next, we examined the physical properties of SFCs hosting different stellar populations. The areas of the SFCs were calculated from the elliptical parameters obtained from SExtractor. Specifically, we used the semimajor and semiminor axes from the SExtractor output to calculate the area of each SFC as $A = \pi ab$, where $a$ and $b$ are the semimajor and semiminor axes, respectively. We find that SFCs containing both EM stars and lower MS stars, as well as SFCs hosting BL stars, generally exhibit the largest areas and highest FUV luminosities. Star-forming clusters containing only EM stars show slightly lower luminosities and smaller areas compared to those with BL stars. Interestingly, SFCs containing only lower MS stars and those containing upper MS stars display comparable areas and luminosities. Nevertheless, one SFC containing only upper MS stars and another containing both upper and lower MS stars exhibit exceptionally low luminosities and small areas. This is puzzling and requires more attention. The SFCs containing two HB stars exhibit area and luminosity distributions similar to those of upper and lower MS SFCs. The distribution of these properties as a function of stellar content is shown in Fig. \ref{fig:leot_stars_prop}.

The larger sizes and higher FUV luminosities of BL and EM-containing SFCs are consistent with relatively recent or prolonged star formation activity. The similarity in properties between upper and lower MS SFCs suggests that FUV emission in these systems is not dominated by a single evolutionary phase. A mild FUV excess in some lower MS SFCs may arise from their location within under-dense regions of the HST field, where incomplete spatial coverage could lead to missing FUV-bright stars.

We consider SFCs hosting FUV sources as extended sources, given their radii of a few parsecs. However, when compared with the HST or MUSE data, we find that most of the SFCs have a single possible FUV-emitting star. This creates uncertainty about whether the FUV source is a single star or a star cluster. This uncertainty may arise from instrumental and analysis-related effects, including UVIT–HST resolution mismatch, spatial variations in UV depth, astrometric uncertainties, and CMD-based stellar classification. In particular, unresolved FUV blends and incompleteness in the resolved stellar catalog may lead to FUV excess in SFCs classified as lower-MS-dominated, while background subtraction and aperture definition can suppress FUV luminosities in compact SFCs. This also creates a large scatter in the properties of SFCs containing a specific type of star, as shown in Fig. \ref{fig:leot_stars_prop}.

This scatter in properties of SFCs makes it difficult to characterize the other outer SFCs. However, by comparing the properties of SFCs without FUV star information with those that are characterized, we can give an idea of what these sources might be. In Table \ref{Table:UV_HST_MUSE}, most of the outer SFCs have luminosities similar to those of MS or HB stars. Only two SFCs have high luminosity similar to BL and EM. This again indicates that most outer SFCs are likely low-mass stars.

\section{Radial profiles of young and old stellar population tracers}\label{Radial trend}
\begin{figure}
    \centering
    \includegraphics[width=0.47\textwidth]{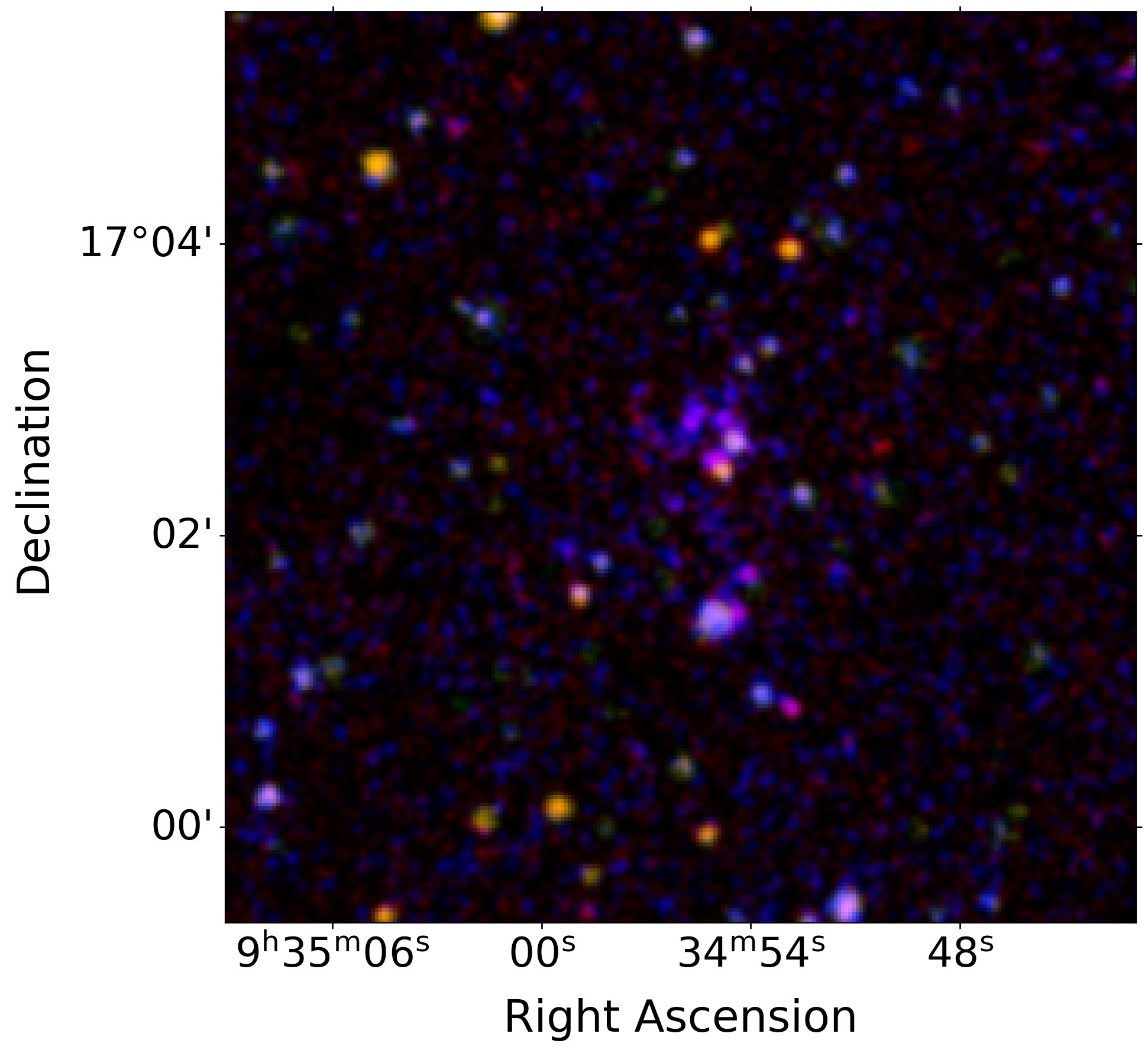}
    \caption{RGB composite image of Leo T using GALEX UV observations. The red channel represents the GALEX NUV emission, the green channel highlights the identified foreground and background sources, and the blue channel represents the GALEX FUV emission.
}
    \label{fig:RGB_image}
\end{figure}

\begin{figure}
   \centering
   \includegraphics[scale=0.35]{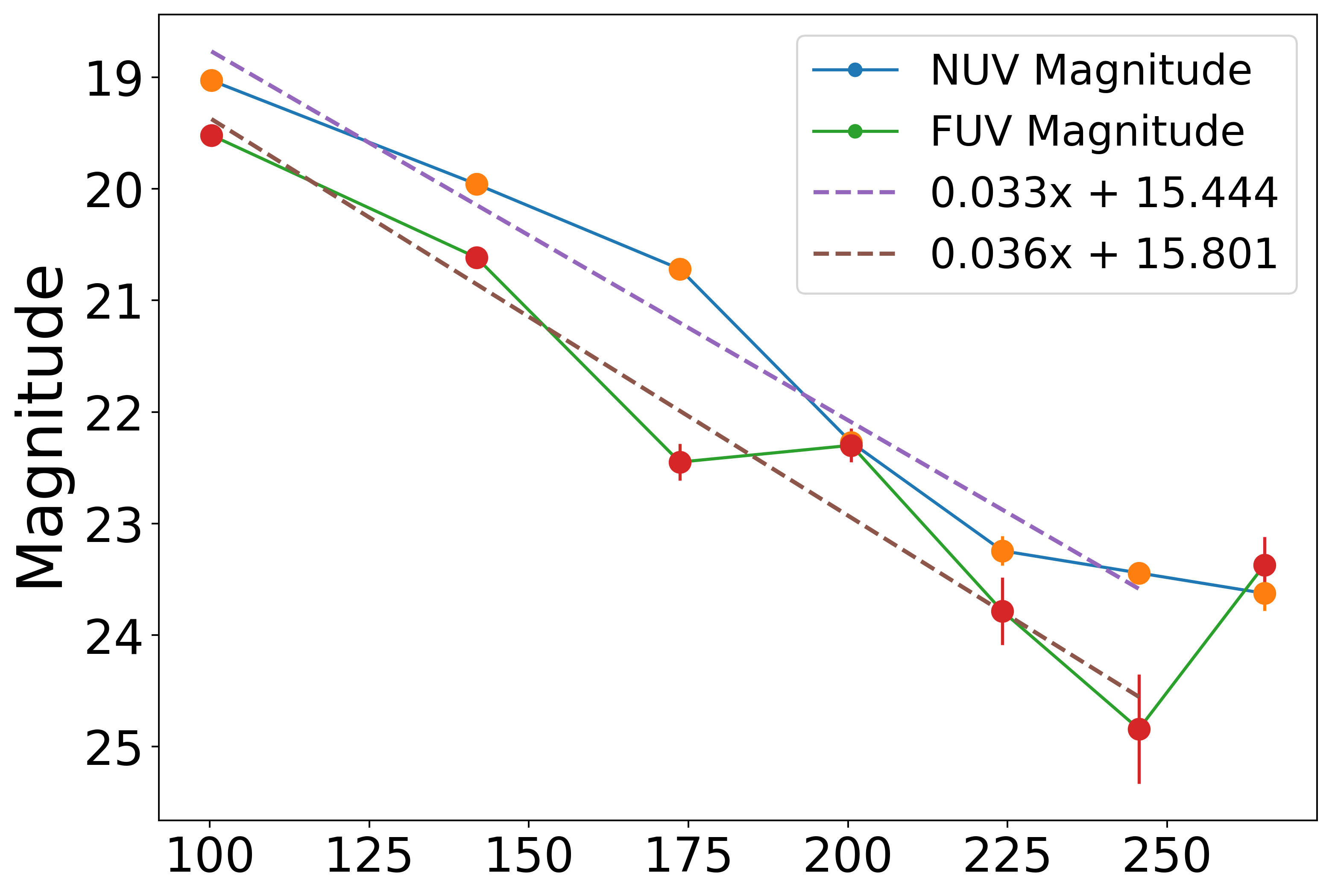}
   \includegraphics[scale=0.35]{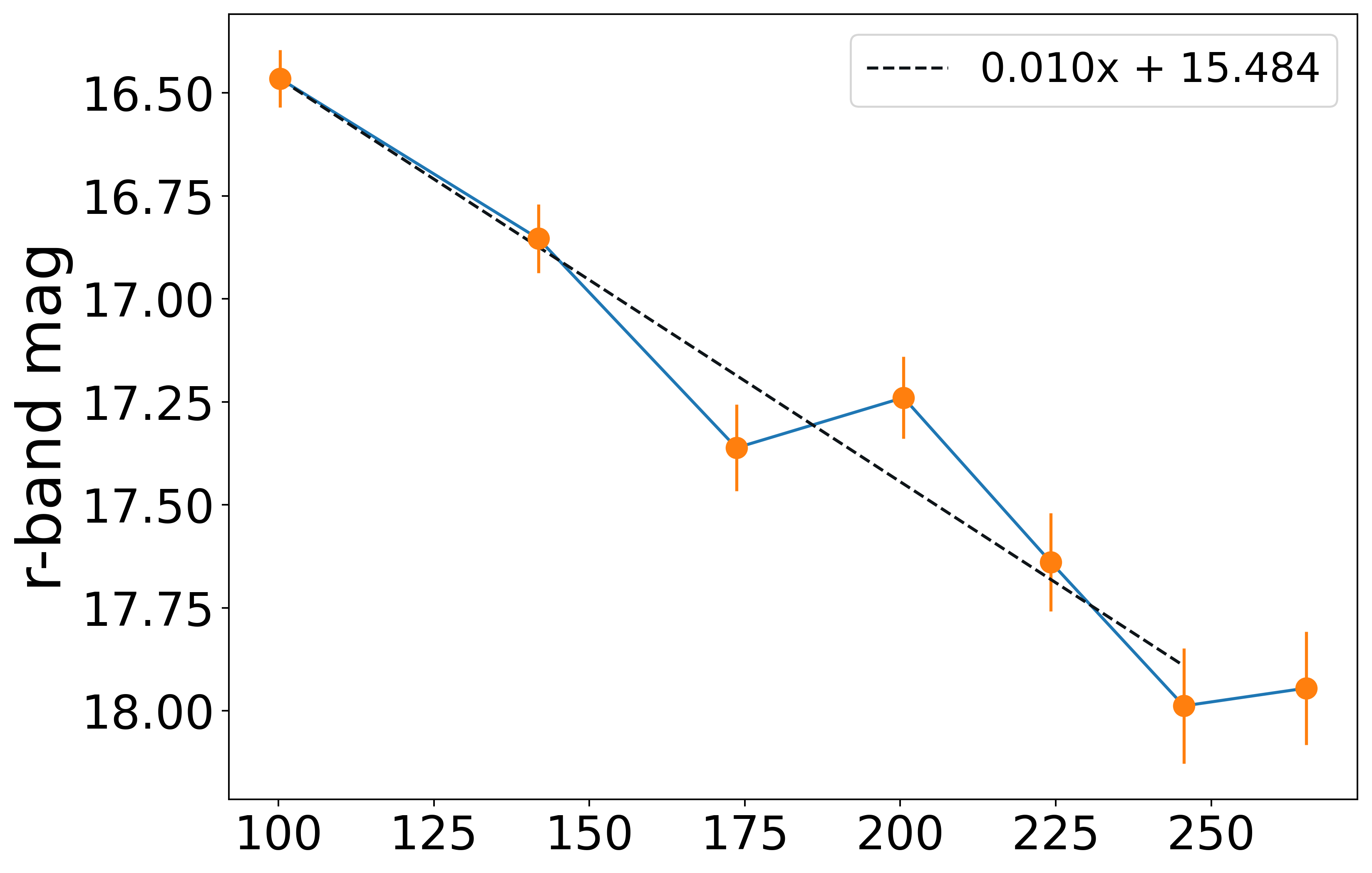}
   \includegraphics[scale=0.35]{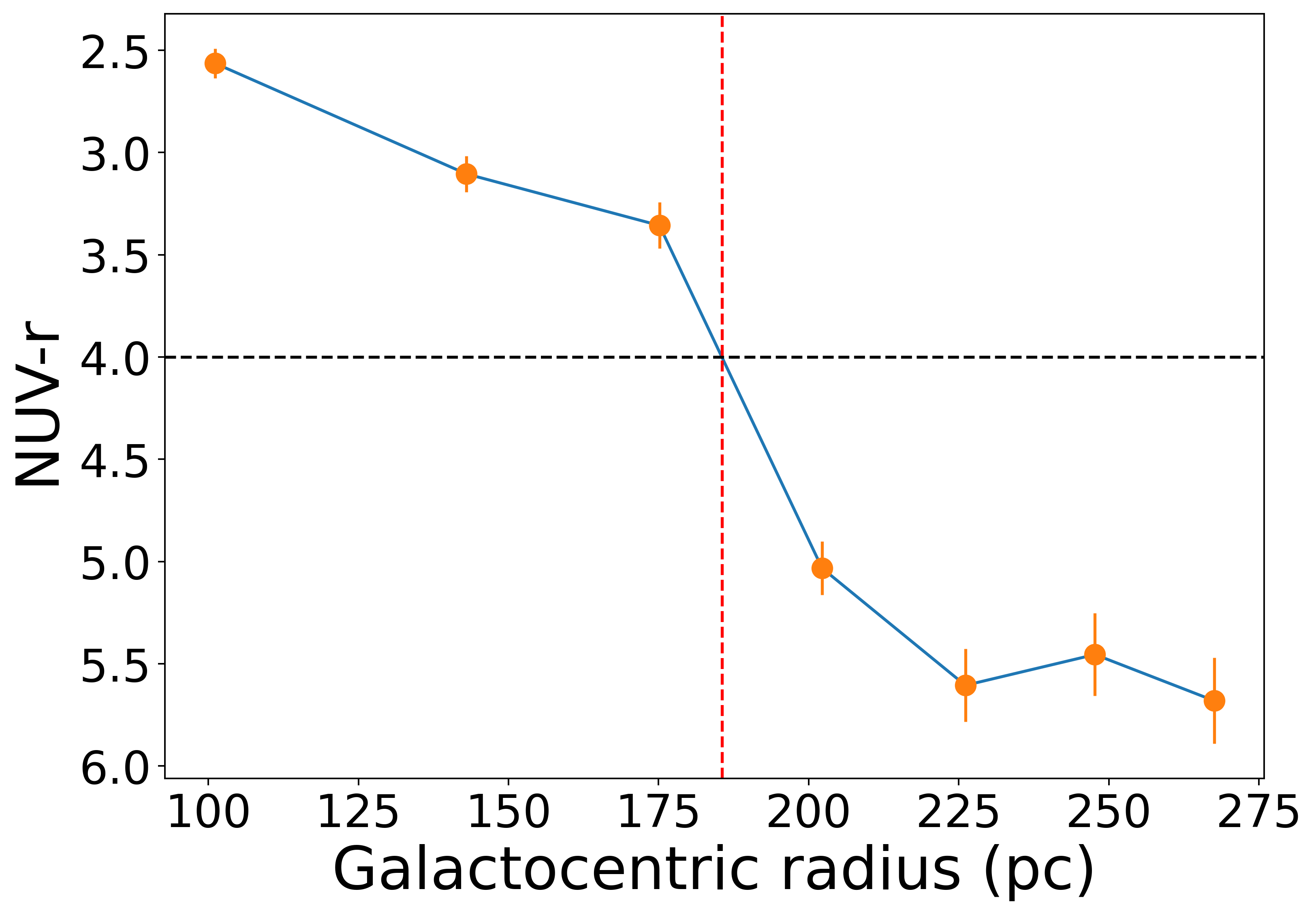}
   \caption{Radial variation of the GALEX FUV and NUV magnitudes (top panel), DECaLS $r$-band magnitude (middle panel), and $\mathrm{NUV}-r$ color (bottom panel). The linear best-fit lines are overlaid on the profiles.
}
              \label{fig:radial_variation}
    \end{figure}

We plotted the radial variation of GALEX FUV, NUV, and DECaLS r-band magnitudes to investigate the spatial distribution of the younger and older stellar populations in Leo T. For that, we convolved GALEX FUV and r-band images to GALEX NUV resolution (5.3$^{\prime \prime}$). Then we considered the total UV emission within concentric annuli centered on the optical center of Leo T and converted the integrated flux in each annulus into a magnitude. We adopted this approach rather than considering only individually resolved sources, as the UV emission may also include unresolved sources due to the sensitivity and spatial-resolution limits of the data. The same procedure was followed for the DECaLS r-band data. This is shown in Fig. \ref{fig:radial_variation}. We used GALEX FUV and NUV data, as UVIT does not have NUV observations for Leo T. The RGB composite image of GALEX FUV and NUV observations is shown in Fig. \ref{fig:RGB_image}. We fitted linear relations to the radial surface-brightness profiles (excluding the last point). Both the FUV and NUV surface brightness profiles exhibit a strong decline with increasing galactocentric radius, fading by approximately 5 mag between 100 and 250 pc, with best-fit slopes of 0.033 and 0.036 mag\,pc$^{-1}$, respectively. The FUV profile shows larger fluctuations and a slightly steeper decline than the NUV profile, suggesting that the youngest stellar populations are more centrally concentrated and that the FUV emission is spatially patchier. The r-band profile follows a similar overall trend but declines by only $\sim$ 2 mag (best fit slope around 0.01 mag\,pc$^{-1})$. This shallower gradient indicates that the older stellar population is more extended than the young UV-emitting component. The slight upturn seen in the outermost radial bin of both the FUV and r bands is likely due to low S/N measured at large radii.

Far-ultraviolet emission can come from both the younger and evolved populations. It is important to determine the type of population that dominates the FUV emission for a galaxy in which 50\% of stars formed 7 Gyr ago. Therefore, we examined the radial variation of $\mathrm{NUV - r}$  over the galaxy. This is shown in the bottom panel of Fig. \ref{fig:radial_variation}. A $\mathrm{NUV - r}$  $<$ 4 indicates that most of the FUV emission is generally dominated by stellar populations younger than $\sim$1 Gyr \citep{Salim_2005, 2007ApJS..173..619K, 2025A&A...696A.118R}. Our radial trend also falls within this value up to 185 pc, indicating that FUV emission over Leo T is dominated by a younger population. \cite{2007ApJS..173..619K} also show that for $\mathrm{NUV - r}$  $>$ 5.5, the emission traces the UV upturn from older populations. Beyond a radius of 225 pc, we observed that $\mathrm{NUV - r}$  $\geq$ 5.5, suggesting a transition to an older stellar population and a diminishing contribution from recent star formation.

Although the $\mathrm{NUV - r}$  color has been extensively used to distinguish between young and old stellar populations in more massive galaxies, its interpretation in an extremely low-mass galaxy such as Leo T should be treated with caution owing to its low metallicity, stochastic SFH, and low stellar mass. Consequently, the radii identified using these empirical $\mathrm{NUV - r}$  thresholds are approximate indicators of changes in the dominant stellar population.

\section{FUV indication of Leo T interaction with MW}\label{interaction effect}

Leo T is the lowest-mass galaxy that has retained its gas. The galaxy may just be falling into the MW for the first time \citep{Rocha_2013} or could be a wide-orbit backsplash dwarf with minimal past interactions with the MW \citep{10.1093/mnras/staa2153, Bla_a_2024}. This interaction may have caused the galaxy to have a gas and stellar center offset \citep{2018A&A...612A..26A}, as well as recent star formation in the extremely low-density and metallicity environment \citep{deJong_2008, Weisz_2012, vaz2023}. Due to this recent star formation, the galaxy emits UV radiation, which helps us constrain and understand the UV-emitting populations.

The FUV observations cover the full extent of the galaxy’s H\,\textsc{i} distribution, enabling us to trace the spatial relationship between FUV-bright stars or clusters and the neutral gas. The presence of an extended H\,\textsc{i} distribution along with a faint gaseous feature to the north is consistent with the action of ram pressure on the system \citep{Bla_a_2024}. Although most SFCs are located close to the optical center and slightly north of it, the southern distribution extends to larger galactocentric distances. Previous studies have also reported asymmetric H\,\textsc{i} morphology, with enhanced gas density toward the south and a compressed edge toward the west \citep{2018A&A...612A..26A,10.1093/mnras/staa2153}. These complex gas and stellar morphologies may reflect a combination of environmental effects, such as ram pressure, and internal dynamical processes \citep{Bla_a_2024}. The observed compression of H\,\textsc{i} could represent an early stage of ram-pressure interaction, in which the remaining gas is compressed before substantial gas removal occurs. Similar leading-edge gas compression and enhanced star formation have been observed in more massive galaxies undergoing ram-pressure stripping \citep{2020ApJ...901...95C}. In the case of Leo T, however, its shallow gravitational potential makes its diffuse gas more susceptible to stripping. Thus, while ram-pressure-induced compression may contribute to the recent star formation, any enhancement is expected to be modest, with continued interaction likely leading to progressive removal of the more diffuse gas. 

The faint northern H\,\textsc{i} extension likely traces stripped or displaced material, where the gas density has dropped below the threshold required for star formation. However, the absence of O- and B-type stars may indicate that such interactions have recently suppressed star formation \citep{vaz2023}. Alternatively, this deficit of massive stars could reflect a truncated high-mass IMF or stochastic sampling of the IMF \citep{Weisz_2012}.

 \section{Star formation rate and star formation efficiency of Leo T}\label{SFR and SFE}

 \subsection{Leo T star formation rate}\label{Leo T Star Formation Rate}

As discussed in Sect. \ref{Properties of UV Star forming Clusters}, H$\alpha$ and UV emission trace star formation over different timescales. H$\alpha$ primarily traces the instantaneous SFR over the past $\sim$10 Myr, whereas UV emission is sensitive to star formation over longer timescales of up to a few hundred Myr \citep{2001ApJ...548..681B, 2004A&A...421..887I, Salim_2007}. From the analysis presented in Sect. \ref{Cross-matching with HST catalog} and Table \ref{Table:UV_HST_MUSE}, we find that several UV-emitting stellar populations in Leo T have ages exceeding 500 Myr. Consequently, the UV luminosity may contain contributions from stars that are significantly older than those responsible for the H$\alpha$ emission, leading to systematically higher UV-derived SFRs. This discrepancy is expected to be particularly pronounced in low-mass galaxies such as Leo T. At very low SFRs, stochastic sampling of the upper end of the IMF can result in a deficit of massive O stars, suppressing the H$\alpha$ luminosity while leaving the UV emission relatively unaffected. In addition, leakage of Lyman continuum photons can further reduce the observed H$\alpha$ flux and lead to an underestimate of the H$\alpha$-derived SFR \citep{Lee_2009}. Therefore, direct comparisons between UV- and H$\alpha$-based SFRs should be interpreted carefully, as the two tracers probe different stellar populations and timescales and may therefore yield different SFR estimates \citep{2015A&A...584A..87C}. In this section, we discuss estimating the upper limit of UV-derived SFR of Leo T and compare it with previous measurements based on other star formation tracers. Any differences between these estimates provide insight into the recent SFH and the stochastic nature of star formation in this extremely low-mass galaxy. 
Leo T has minimal H$\alpha$ emission, and within the small MUSE FoV, the upper limit of the H$\alpha$ SFR density $\Sigma_{\text{SFR}}$ is less than $10^{-5} \, \text{M}_{\odot} \, \text{yr}^{-1} \, \text{kpc}^{-2}$ \citep{vaz2023}. At extremely low H$\alpha$ SFRs, as in Leo T, the standard H$\alpha$-SFR calibration may become uncertain, and the H$\alpha$-derived SFR may differ substantially from SFR estimates based on tracers sensitive to longer-lived stellar populations. Hence, we calculated the $\Sigma (\rm SFR_{FUV})$ within the effective radius \citep[115 pc;][]{10.1111/j.1365-2966.2010.16753.x} and out to the farthest detected FUV SFC (325 pc). These yield values of $5 \times 10^{-5}\mathrm{M_\odot yr}^{-1} \mathrm{kpc}^{-2}$ and $1.3 \times 10^{-4}\mathrm{M_\odot yr}^{-1}\mathrm{kpc}^{-2}$, respectively. The corresponding $\Sigma (\rm SFR_{FUV})$ values were derived by converting the integrated FUV luminosities within these radii to SFRs using the calibration given in Equation~\ref{eq1}, and subsequently dividing the resulting SFRs by the corresponding projected areas.
 The $\Sigma (\rm SFR_{FUV})$ within the effective radius is five times larger than the H$\alpha$ upper limit \citep{vaz2023}. Interestingly, the $\Sigma (\rm SFR_{FUV})$ for the full extent is larger than the value within the effective radius. However, as shown in Sect. \ref{Radial trend}, the UV emission at large radii is increasingly dominated by evolved stellar populations. Consequently, the integrated FUV luminosity may be contaminated, causing the inferred galaxy-wide $\Sigma (\rm SFR_{FUV})$ to be overestimated. Even with this possible contamination, we find that the upper limit of $\Sigma (\rm SFR_{FUV})$ is lower than $\Sigma_{\text{SFR}} = 10^{-3} \, \text{M}_{\odot} \, \text{yr}^{-1} \, \text{kpc}^{-2}$, which was derived using SFHs from stellar population models \citep{Weisz_2012}. Overall, the observational $\Sigma (\rm SFR_{FUV})$ for Leo T is at the lower end of those found in extremely metal-poor dwarfs \citep{2016ApJ...820..109F}.

 The individual SFCs exhibit larger $\Sigma (\rm SFR_{FUV})$ values than the integrated average, as expected given their compact sizes. Although no gradient appears in the sizes of SFCs as we go from a higher-density to a lower-density region, the $\Sigma (\rm SFR_{FUV})$ is generally higher for SFCs located closer to the optical center and lower in the outskirts. This pattern is consistent with the expectation that star formation proceeds more efficiently in regions of higher stellar and gas density. 
 
\subsection{Correlation with CO clouds and star formation efficiency}\label{Correlation with molecular hydrogen and star formation efficiency}
\begin{figure}
   \centering
   \includegraphics[scale=0.31]{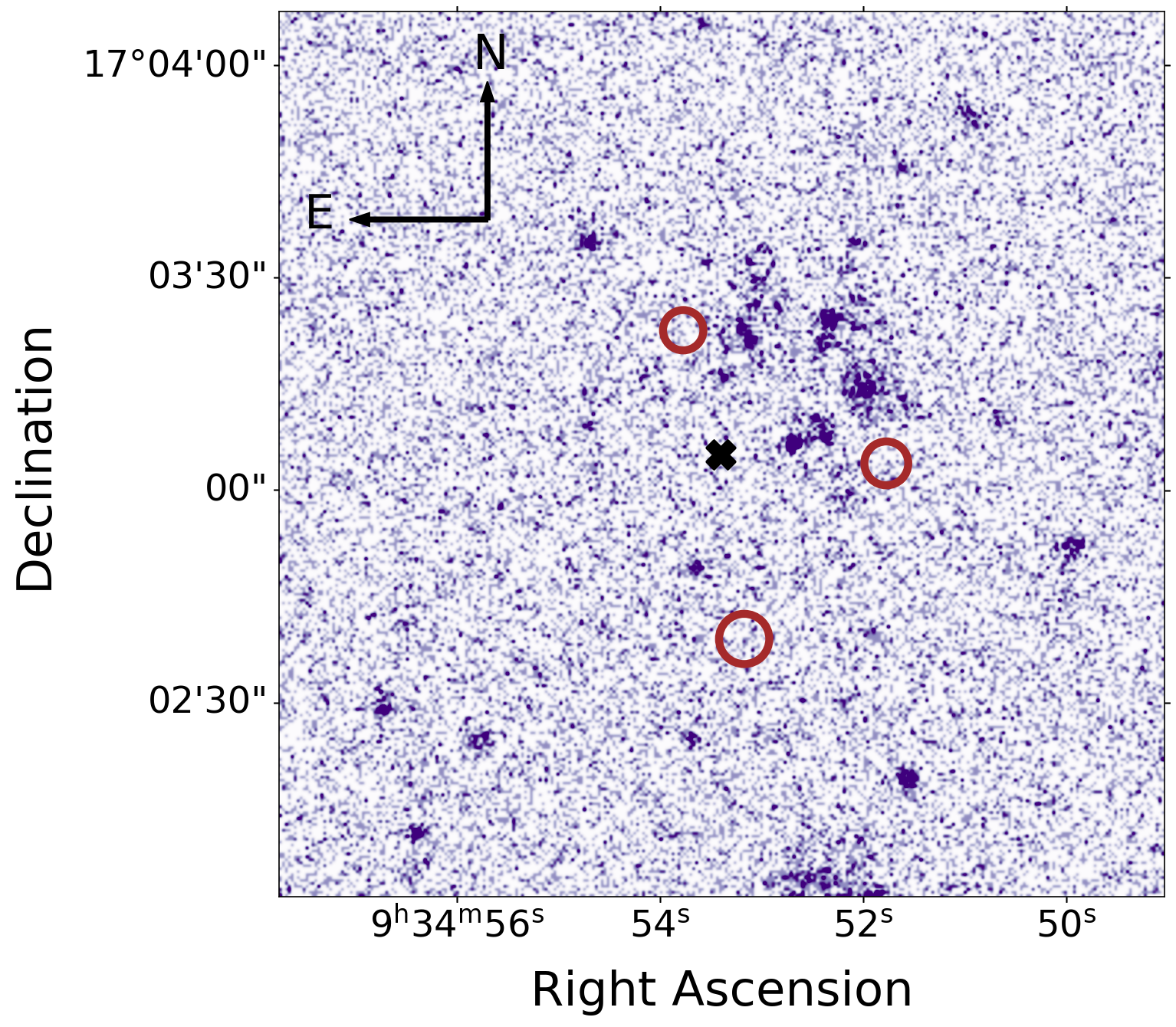}
   \caption{UVIT FUV image of Leo T. The brown circles indicate the positions and sizes of the CO clouds detected by \citet{2025A&A...699L..11V}. The aperture radii were adopted from their analysis. The FUV fluxes measured within these apertures were used to estimate the upper limits on the SFRs and SFEs associated with the molecular clouds.}
              \label{fig:CO-SFE}
    \end{figure}

Recently, \citet{2025A&A...699L..11V} reported the first detection of CO(1–0) emission in Leo T, identifying three compact molecular clouds that are notably spatially offset ($\sim 60$ pc) from the main stellar population centers. The locations of these regions are shown as brown circles in Fig. \ref{fig:CO-SFE}. No SFCs are detected within these CO-bright regions. We measured the FUV emission within the CO apertures and derived the corresponding upper limit on the SFRs. The upper limit on the SFRs, molecular gas masses, and resulting SFEs are listed in Table \ref{Table:CO}. The mean upper limits for $\mathrm{log(SFR / M_\odot yr^{-1}}$) and $\mathrm {log(SFE / yr^{-1}}$) are -8.76 and -12.48, respectively, indicating either extremely low star formation activity and low SFE in all three regions, or that star formation has already faded below FUV detectability. Another possibility is that the CO-bright region is expected to have significant internal extinction, rendering any UV emission invisible. Therefore, the absence of detected UV SFCs within the CO-bright regions does not, by itself, imply an absence of recent star formation in these regions. 

The low SFR and SFE are consistent with stochastic star formation, where gas can remain molecular for long periods without forming stars. This results in longer gas depletion times and the inefficient collapse of molecular clouds. Among the three clouds, the southern CO region has the highest SFE, despite being located farther from the SFCs and the galaxy's optical center. However, because this cloud lies within a dense H\,\textsc{i} region \citep{2025A&A...699L..11V}, it could be influenced by external processes, making it slightly distinct from the central areas. 

\cite{2025A&A...699L..11V} also report that the northern cloud near the optical center has very high v$_{los}$ and $\mathrm{\Delta V_{CO-HI}}$=57.7 $\pm$ 0.7 $\mathrm{km^{-1}}$. This high-velocity offset is associated with the expulsion of molecular gas, which may explain why this cloud exhibits the lowest SFE compared to gravitationally bound clouds.

% Please add the following required packages to your document preamble:
% \usepackage{graphicx}
\begin{table}[]
\centering
\caption{Physical quantities of the CO clouds in Leo T.}
\label{Table:CO}
\resizebox{\columnwidth}{!}{%
\begin{tabular}{ccccccc}

\hline\\

Cloud  & $\mathrm{r_{CO}}$       & $\mathrm{M_{mol}}$          & $\mathrm{SFR_{FUV}}$                  & SFE                 \\

Position   & [arcsec]       & [$\mathrm{10^3 M_{\odot}}$] & [$\mathrm{10^{-9} M_{\odot}yr^{-1}}$] & [$\mathrm{10^{-13}yr^{-1}}$] &          \\

\hline\\

N& 5.7 $\pm$ 0.25 & 4.6 $\pm$ 1.4      & $<$0.96 $\pm$ 0.08              & $<$2.08 $\pm$ 0.66           \\
W & 6.2 $\pm$ 0.8  & 5.0 $\pm$ 1.6      & $<$1.53 $\pm$ 0.08              & $<$3.06 $\pm$ 0.99             \\
S & 7.1 $\pm$ 0.23 & 4.9 $\pm$ 1.5      & $<$2.74 $\pm$ 0.08              & $<$5.59 $\pm$ 1.71         \\

\hline
\end{tabular}%
}
\end{table}

\section{Conclusions}\label{sec Conclusions}

 In this paper, we presented a UVIT FUV study of the MW satellite galaxy Leo T, which is one of the lowest-mass, gas-rich MW satellites that show signs of recent star formation. Using UVIT's high resolution and large FoV, we extracted the SFCs and examined their properties in this galaxy and compared them with HST and MUSE data to identify the possible UV-emitting stellar populations and their stellar parameters. Our main results of this work are summarized below:

1. The distribution of SFCs in Leo T is extended beyond 300 pc. However, 70~\% of SFCs are within the half-light radius (R$_{1/2}$=115pc).

2. The FUV luminosity of the SFCs is less than the FUV luminosity of massive O and B stars. This confirms previous results, indicating no massive O and B stars in Leo T. This finding either constrains recent star formation to at least 30 Myr ago or suggests stochastic star formation in the galaxy.

3. The $\Sigma (\rm SFR_{FUV})$ of SFCs of Leo T is less than the $\Sigma (\rm SFR_{FUV})$ of SFCs of the SFDGs.

4. We cross-matched SFCs with the HST and MUSE stellar population catalog and find possible UV-emitting stars. The FUV-emitting stars in Leo T are upper MS, lower MS, BL, HB, and EM stars.
Five out of six FUV-emitting MUSE stars are extremely metal-poor ([Fe/H]$\leq$-3). 

5. Upper MS and EM stars are concentrated near the optical radius. Whereas lower MS and BL stars are spread even slightly farther from the optical center.

6. We find SFCs with BL and EM stars generally have larger radii and luminosities compared to the other SFCs. Interestingly, SFCs with upper and lower MS have similar properties, which suggests some uncertainty due to  UVIT–HST resolution mismatch, spatial variations in UV depth, astrometric uncertainties, and CMD-based stellar classification.

7. Leo T has an ($\mathrm{NUV - r}$) color profile, which indicates that the UV emission is dominated by stellar populations younger than 1 Gyr out to a radius of $\sim$ 185 pc. Beyond this radius, emission is dominated by older populations.

8. The SFCs are concentrated around the optical center and have extended farther south of the optical radius, suggesting a directional interaction with the MW’s halo.

9. Leo T has a very low upper limit on $\Sigma (\rm SFR_{FUV})$, which is below 10$^{-4}$M$_{\odot}$yr$^{-1}$pc$^{-2}$, placing it at the lower end of extremely metal-poor dwarfs. 

10. When compared with CO clouds from \cite{2025A&A...699L..11V}, we do not see any bright FUV emission associated with it. Hence, the SFE limit is also very low. However, the southern CO region exhibits a higher SFE compared to the other regions, indicating that it may be influenced by external processes.

Overall, our analysis provides insights into the stellar populations responsible for the FUV emission in the extremely low-mass galaxy Leo T using spatially resolved FUV observations, while also highlighting the limitations of individual star-formation tracers in the extremely low-SFR regime. The large FoV and relatively high spatial resolution of UVIT enable us to investigate the spatial distribution of the FUV emission across Leo T, providing information that was not accessible from previous high-resolution observations with smaller FoVs. However, the interpretation of the FUV emission remains subject to uncertainties associated with stochastic sampling of the stellar IMF and the small number of young stars. Future observations with deeper and higher-resolution UV imaging, combined with improved modeling of stochastic star formation and optical spectroscopy, will provide tighter constraints on the spatial distribution and recent SFH of Leo T. Combining these datasets with resolved stellar-population modeling and simulations of Leo T's interaction with the MW will further help clarify the nature of the stellar populations producing the UV emission and the role of environmental effects in shaping recent star formation in such an extremely low-mass, gas-rich galaxy.

\begin{acknowledgements}
We thank the anonymous referee for the constructive comments and suggestions, which helped us improve the manuscript. We thank Prof. Dan Weisz for kindly providing the HST catalog upon our request. MD acknowledges the support of the Department of Science and Technology (DST) grant DST/WIDUSHI-A/PM/PM/2023/25(G) for this research. This publication uses data from UVIT onboard the AstroSat mission of the Indian Space Research Organization (ISRO) and is archived at the Indian Space Science Data Centre (ISSDC). We are grateful to the members of various teams for supporting the project from the early stages of design to launch and observations in orbit. This work uses observations made with ESO Telescopes at the La Silla Paranal Observatory under program IDs 0100.D-0807, 0101.D-0300, 0102.D-0372, and 0103.D-0705.
\end{acknowledgements}

% for the bibliography, at the end
\bibliographystyle{aa} % style aa.bst
\bibliography{References}

\end{document}